\documentclass{article}
\usepackage{iclr2027_conference,times}
\usepackage{amsmath,amssymb,booktabs,array}
\usepackage{longtable}
\usepackage{xcolor}
\usepackage{colortbl}
\usepackage{graphicx}
\usepackage{tikz} 

\definecolor{rankgold}{HTML}{D5A526}
\definecolor{ranksilver}{HTML}{A6AFBA}
\newcommand{\rankcrown}[1]{\hspace{0.08em}\tikz[x=0.65em,y=0.65em,baseline=0pt]{%
  \path[fill=#1,draw=#1!65!black,line width=0.18pt,line join=round]
    (0.08,0.08)--(0,0.66)--(0.25,0.43)--(0.45,0.85)--(0.65,0.43)--(0.9,0.66)--(0.82,0.08)--cycle;
  \draw[#1!65!black,line width=0.2pt] (0.1,0.2)--(0.8,0.2);
}}
\newcommand{\goldcrown}{\rankcrown{rankgold}}
\newcommand{\silvercrown}{\rankcrown{ranksilver}}
\newcommand{\nocrown}{\phantom{\goldcrown}}

\graphicspath{{figures/}}
\usepackage{float}
\usepackage{flafter}
\usepackage{placeins} 
\usepackage{titletoc}
\usepackage{hyperref}
\usepackage{url}
\definecolor{softblue}{HTML}{464BB4}
\title{MolP2SBench: Do LLMs Really Reason from Molecular Properties to Structure?}
\author{%
\makebox[\dimexpr\textwidth-2\tabcolsep\relax][c]{%
\parbox[t]{6.5in}{\centering\normalfont\fontsize{10}{12}\selectfont
\textbf{Yuhang Xie\textsuperscript{1,*}, Zhiyi Shi\textsuperscript{2,*}, Zeyi Zhang\textsuperscript{3}, Zhisheng Qi\textsuperscript{4}, Yushun Dong\textsuperscript{5}, Yu Wang\textsuperscript{4}, Xiongye Xiao\textsuperscript{1,\ensuremath{\dagger}}}\\[0.7em]
\textsuperscript{1}University of Tennessee, Knoxville\quad
\textsuperscript{2}University of Illinois at Urbana-Champaign\\[0.2em]
\textsuperscript{3}University of California, Berkeley\quad
\textsuperscript{4}University of Georgia\quad
\textsuperscript{5}Florida State University
}}}
\iclrfinalcopy
\begin{document}
\setlength{\textfloatsep}{\baselineskip}
\maketitle
\begingroup
\renewcommand{\thefootnote}{\fnsymbol{footnote}}
\footnotetext[1]{Equal contribution.\quad\textsuperscript{\ensuremath{\dagger}}Corresponding author: Xiongye Xiao (\texttt{xxiao9@utk.edu}).}
\endgroup
\lhead{Preprint}

\begin{abstract}
Large language models (LLMs) are increasingly used in molecular discovery and structure elucidation, where decisions depend on physical observations. Existing evaluations primarily focus on end-to-end molecular reconstruction, where correct predictions do not necessarily indicate that these results are grounded in the supplied properties. To address this, we introduce MolP2SBench, a benchmark for evaluating molecular property-to-structure (P2S) reasoning in LLMs. Each task presents a masked molecular scaffold with given properties, requiring the model to select the missing structure from plausible candidates. To assess whether predictions genuinely rely on the provided properties, we test how models respond when properties are shuffled or replaced, thereby evaluating their reliance on correct property-structure pairing and their responsiveness to counterfactual property changes.
We also establish a channel-wise leaderboard using Property-Grounded Recovery (PGR), which gives equal weight to source and counterfactual recovery. In our evaluation of seven models, we find that LLMs can leverage physical properties for molecular structure prediction, but their performance varies substantially across property types. More importantly, higher accuracy with a given property does not necessarily imply greater responsiveness to that property. For example, a model may benefit more from $^1$H NMR than from $^{13}$C NMR, yet be less likely to change its prediction when the $^1$H data is replaced. Interestingly, LLM performance does not necessarily reflect the relative difficulty of these properties for human chemists. For instance, although UV–Vis spectra provide less direct structural information than NMR spectra, Astra achieves higher accuracy with UV–Vis data than with NMR peaks. In summary, MolP2SBench provides a framework for studying how LLMs extract and use information from different physical properties, revealing performance patterns that may differ from expectations based on conventional chemical intuition and offering a foundation for developing more reliably property-grounded models for molecular and quantum-chemical inference. The code and dataset are available at \href{https://github.com/DUAL-Xiao/MolP2SBench}{https://github.com/DUAL-Xiao/MolP2SBench}.

\end{abstract}

\section{Introduction}
\label{sec:introduction}
LLMs are increasingly used in molecular science, including molecular generation~\citep{cai2026mollangbench}, molecular discovery~\citep{ye2025drugassist}, and structure elucidation~\citep{guo2024molpuzzle,su2026spectrallm,noh2026iragent}. In many of these settings, decisions depend on more than recognizing molecular motifs; they require relating physical observations, such as spectra, electronic properties, or molecular geometry, to competing structural hypotheses. Different property channels encode different structural information, and models may not exploit them equally well. Thus, a key question is not simply whether an LLM can produce the correct molecular structure, but whether its decision actually responds to the physical evidence provided.

\begin{figure}[!t]
\centering
\includegraphics[width=\linewidth]{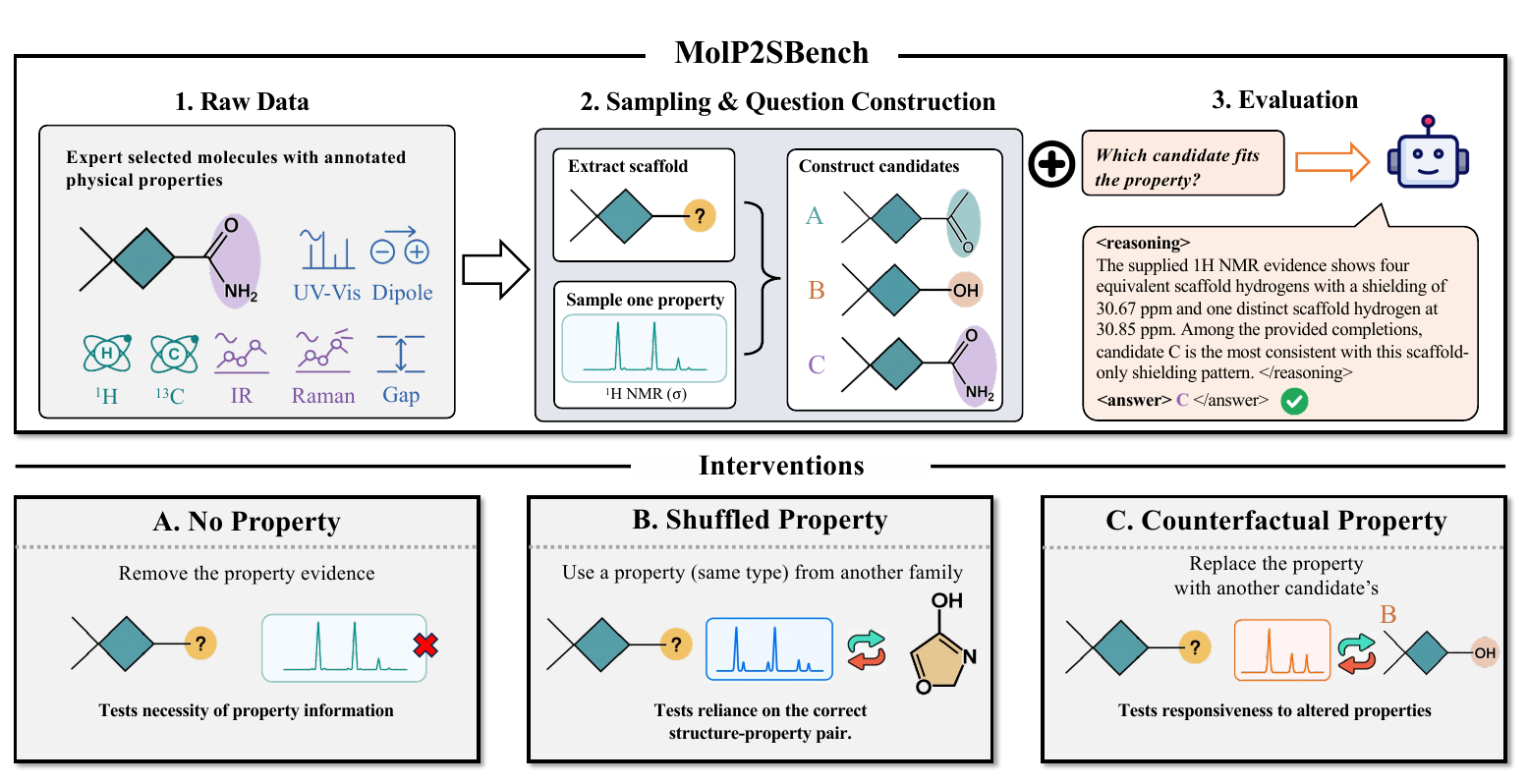}
\caption{\textbf{MolP2SBench: a fixed molecular choice under changing property inputs.} \textbf{Top:} a masked scaffold, candidate SMILES, and physical observations define a local-completion task; the model selects a candidate and provides a short justification. The illustrative $^1$H input contains calculated shieldings and counts for retained scaffold nuclei. Molecular drawings visualize the text inputs, and the response illustrates the output format. \textbf{Bottom:} structure-only input, cross-task shuffling, and same-scaffold counterfactual (CF) property substitution test the benefit of property inputs, pairing dependence, and whether the choice follows replacement property information. The scaffold and candidates remain fixed throughout.}
\label{fig:task_overview}
\label{fig:overview}
\end{figure}

Existing evaluations~\citep{liu2025fgbench,bushuiev2024massspecgym,han2026molquest} do not directly answer this question. End-to-end molecular reconstruction entangles evidence interpretation with structural search and output generation: a model may reach the correct answer because of structural priors, even if the supplied property evidence plays little role in its decision~\citep{geirhos2020shortcut}. Aggregate accuracy does not reveal which property channels are actually useful to the model, nor whether the model would change its prediction if the evidence changed; gains under property-conditioned input therefore do not by themselves establish evidence use~\citep{feng2019misleading}.

To systematically evaluate how well models use the supplied physical properties, we formulate property-to-structure prediction as a multiple-choice masked completion
task (Figure~\ref{fig:task_overview}). Each task presents a masked molecular scaffold with given properties, requiring the
model to select the missing structure from plausible candidates. To evaluate how well models use the supplied properties, we apply three controlled intervention schemes. Removing them tests whether the supplied properties confer any benefit beyond the scaffold alone; shuffling them across tasks measures the model's reliance on correct property-structure correspondence; and replacing them with counterfactual (CF) properties from an alternative candidate within the same family assesses the model's responsiveness: whether its selection dynamically adapts to the newly supplied evidence~\citep{gardner2020contrast}. Collectively, these comparisons address not merely whether a model is correct, but whether its decision tracks the physical observations it was given.

Our results show that while LLMs can leverage physical properties for molecular structure prediction, their reliance on this data varies substantially across models and property types. Reliance and responsiveness also show different observed patterns across property channels. For example, Sol has a larger accuracy gain from correctly paired atomwise $^1$H NMR than from $^{13}$C peaks, yet a lower conditional flip rate toward the corresponding CF candidate when the atomwise $^1$H properties are substituted. Interestingly, model performance does not consistently reflect the relative difficulty of these properties for human chemists. For instance, although UV–Vis spectra provide less direct structural mapping than NMR, models such as Astra achieve higher accuracy with UV–Vis data than with familiar NMR peaks. Beyond these interventions, combined-input analyses reveal model-specific gains and losses: extending the strongest property bundle to all properties improves recovery for Sol and Astra but reduces it for Luna and Opus. Furthermore, in paired NMR comparisons, net accuracy gains often hide significant shifts in model behaviors, as higher-recovery models still fail on sizeable subsets of tasks successfully solved by weaker models. Case studies show that while Astra demonstrates a superior ability to utilize local chemical structures for NMR interpretation, all evaluated models systematically err when global molecular dipoles contradict local polarity heuristics.

In summary, whether a model selects the correct structure, whether its decision is genuinely grounded in the supplied evidence, and whether its stated justification is consistent with the evidence are distinct questions. MolP2SBench disentangles these dimensions through controlled intervention schemes and targeted case studies, exposing variations in property utilization that traditional accuracy metrics fail to reflect. Our contributions are fourfold:

\noindent\textbf{1. A channel-wise recovery leaderboard.} We establish a leaderboard using Property-Grounded Recovery (PGR), which balances source recovery with recovery under candidate-targeted counterfactual property substitutions. By averaging counterfactual recovery within each family prior to cross-family aggregation, PGR ensures equal weighting regardless of varying candidate counts. Furthermore, reporting scores independently across property channels reveals model-specific strengths without collapsing heterogeneous properties into a single aggregate ranking.

\noindent\textbf{2. Disentangling reliance and responsiveness.} We introduce two complementary metrics to characterize distinct capabilities in how models utilize supplied molecular properties. To measure static reliance, Reliance Margin (RM) quantifies the performance change when the correct property–task pairing is disrupted by shuffling. To evaluate dynamic responsiveness, Counterfactual Flip Rate (CFR) measures whether predictions follow targeted substitutions of the supplied property among cases that are initially solved correctly. The observed channel rankings differ between the two metrics, motivating separate evaluation of reliance and responsiveness.

\noindent\textbf{3. Substitution-ready benchmark construction.} We construct MolP2SBench with 136 distinct task families spanning 11 molecular-property channels, yielding 1,496 property-conditioned source tasks paired with matched shuffled controls and candidate-targeted counterfactual variants. Additionally, we analyze within-family property separability and verify that our benchmark produces robust metric estimates and stable model rankings under substantial subsampling.

\noindent\textbf{4. Beyond aggregate accuracy.} We show that recovery depends strongly on the supplied properties, and that a higher-recovery model does not simply solve a superset of a weaker model's tasks: across paired comparisons, net gains span 11.0 to 32.4 pp while 40.4\% to 45.6\% of tasks change outcome in one direction or the other. Model explanations can also conflict with candidate-level physical values, with all four closed-weight models failing more often when molecular dipoles contradict a local carbonyl-based polarity heuristic.

\section{Related Work}
\label{sec:related-main}

Table~\ref{tab:related_work_comparison} compares MolP2SBench with related molecular benchmarks.  We provide a more detailed discussion of the related work in Appendix~\ref{sec:related}.

\begin{table}[H]
\centering
\setlength{\abovecaptionskip}{0pt}
\caption{\textbf{Related work comparison.} We compare benchmark support for resolving structure from input properties. $\checkmark$ marks an evaluated feature; $\times$ means it was not reported or not part of the task.}
\label{tab:related_work_comparison}
\fontsize{9}{10}\selectfont
\setlength{\tabcolsep}{1.5pt}
\renewcommand{\arraystretch}{1.0}
\begin{tabular*}{\textwidth}{@{\extracolsep{\fill}}lccccc@{}}
\toprule
Benchmark & {\shortstack{\strut Local editing/\\[-2pt]\strut completion}} & {\shortstack{\strut Masked-part\\[-2pt]\strut reconstruction}} & {\shortstack{\strut Structural\\[-2pt]\strut distractors}} & {\shortstack{\strut Property-driven\\[-2pt]\strut answer switch}} & {\shortstack{\strut Explanation\\[-2pt]\strut checks}} \\
\midrule
FGBench~\citep{liu2025fgbench} & $\times$ & $\times$ & $\times$ & $\times$ & $\checkmark$ \\
ChemCoTBench~\citep{li2025chemcot} & $\checkmark$ & $\times$ & $\times$ & $\times$ & $\times$ \\
MolLangBench~\citep{cai2026mollangbench} & $\checkmark$ & $\times$ & $\times$ & $\times$ & $\times$ \\
ChemIQ~\citep{runcie2026chemiq} & $\times$ & $\times$ & $\times$ & $\times$ & $\checkmark$ \\
SpectrumBench~\citep{yang2026spectrumworld} & $\times$ & $\times$ & $\checkmark$ & $\times$ & $\times$ \\
MolecularIQ~\citep{bartmann2026moleculariq} & $\times$ & $\times$ & $\times$ & $\times$ & $\checkmark$ \\
ChemDIRT~\citep{inae2026chemdirt} & $\times$ & $\times$ & $\checkmark$ & $\times$ & $\times$ \\
ChemCoTBench V2~\citep{guo2026chemcotv2} & $\checkmark$ & $\times$ & $\times$ & $\times$ & $\checkmark$ \\
\midrule
\textbf{MolP2SBench (ours)} & $\checkmark$ & $\checkmark$ & $\checkmark$ & $\checkmark$ & $\checkmark$ \\
\bottomrule
\end{tabular*}
\end{table}

\section{Benchmark Design}
\label{sec:benchmark}
MolP2SBench formulates property-to-structure prediction as a multiple-choice masked completion task to systematically evaluate how well models use the supplied physical properties. Each task's input is carefully constructed around three core elements: (i) a \textbf{molecular scaffold} where property-critical fragments are masked; (ii) \textbf{physical properties} chosen for their high sensitivity to structural changes; (iii) a \textbf{candidate family} list containing a unique ground-truth completion. Through controlled intervention on the provided properties, this design explicitly isolates and evaluates two key dimensions of model capability beyond pure reconstruction: \textbf{reliance} on correctly paired properties; \textbf{responsiveness} to counterfactual properties. Figure~\ref{fig:task_overview} summarizes the model-visible inputs and the intervention scheme.

\subsection{Problem Formulation}
\label{sec:task}
\noindent\textbf{Masking} isolates local structural inference from open-ended molecular generation. Reconstructing an entire molecule usually combines evidence interpretation with structural search and output generation. Evaluating such outputs is also ambiguous when a generated structure differs from the reference but exhibits similar properties. We therefore mask property-critical fragments while preserving the remaining scaffold and boundary bond orders. This focuses the task on a localized structural difference within a fixed molecular context.

\noindent\textbf{Property Selection.} We select physical properties from three groups that characterize different aspects of molecular structure: electronic properties comprising dipole magnitude, the HOMO--LUMO gap, and UV--Vis observations; vibrational properties comprising IR and Raman spectra; and NMR properties comprising $^1$H and $^{13}$C observations. These groups characterize electronic structure, bond vibrations, and local nuclear environments, respectively. We obtain \textbf{11 property channels} by evaluating each electronic and vibrational property separately and using three representations for each NMR type: peaks-only, integration-augmented, and atomwise for $^1$H; peaks-only, degeneracy-augmented, and atomwise for $^{13}$C. NMR inputs provide calculated shielding values for nuclei in the visible scaffold. Representation details appear in Appendix~\ref{app:properties}.

\noindent\textbf{Candidate families} make property interventions directly comparable.
\label{sec:distinguishability}
Each task provides two to four chemically valid options compatible with the scaffold and its attachment constraints. Construction combines three stages: (i) scaffold matching groups molecules with an identical canonical scaffold after masking; (ii) family assembly selects structurally distinct completions, maintains molecule-disjoint tasks, and aligns their electronic, vibrational, and NMR properties; (iii) property-specific filtering excludes a family if any verifiable candidate pair falls below its channel-specific separation threshold. 
    
The final benchmark contains 136 task families and 394 distinct candidate structures. Its 11 property representations provide 1,496 source-property instances, matched shuffled-property controls, and 2,766 eligible within-family CF substitutions. Across 1,000 subsamples of 100 families, aggregate rankings for the diagnostic metrics RM and CFR are retained in 97.7--99.3\% of draws (Appendix~\ref{app:family-subsampling}). Channel-wise PGR yields mean rank correlations of 0.940--0.999 and mean absolute score deviations of 0.91--1.34 pp (Appendix~\ref{app:pgr-subsampling}).
This organization supports controlled comparisons of property reliance and responsiveness with fixed structural alternatives. Construction and coverage details appear in Appendix~\ref{app:release}. 

\subsection{Task Construction}\label{sec:protocol}
Let $q_i$, for $i=1,\ldots,N$, denote a task with visible scaffold $G_i^{\mathrm{obs}}$ and ordered candidate family $\mathcal C_i=\{G_{i0},\ldots,G_{i,n_i-1}\}$, where $n_i$ ranges from two to four. The source candidate $G_{i s_i}$ is the molecule from which the task is constructed, with source index $s_i$. For property configuration $p$, let $\mathbf P_{ij}^{(p)}$ denote candidate $j$'s aligned observations. We keep the scaffold and candidate order fixed while varying the supplied properties to construct two types of evaluation tasks. The model receives one property input, not a lookup table of candidate properties, and returns a zero-based candidate index with a concise justification.

\noindent\textbf{Reliance tasks} evaluate whether successful selection depends on correctly paired properties. We compare two conditions: \textbf{source properties} $\mathbf P_{i s_i}^{(p)}$, the observations associated with the source candidate; and \textbf{shuffled properties} $\mathbf P_{\pi(i),s_{\pi(i)}}^{(p)}$, which break the task--property correspondence while preserving the property channel and input format. Here, $\pi$ assigns each evaluated task a different source from a fixed reference property pool; donor assignments are held constant across channels (Appendix~\ref{app:release}). Both conditions are evaluated against $s_i$, and their accuracy difference measures the benefit of correct property pairing. We denote the corresponding predictions by $\widehat s_{i,\mathrm{src}}^{(p)}$ and $\widehat s_{i,\mathrm{shuf}}^{(p)}$. We additionally include a \textbf{no properties} baseline $P_0=\varnothing$, shared across configurations, to measure the benefit of supplying properties beyond the structural input alone. Its prediction, $\widehat s_{i,0}$, is also evaluated against $s_i$.

\noindent\textbf{Counterfactual tasks} evaluate whether the model's selection is sensitive to counterfactual changes in the provided properties. We replace $\mathbf P_{i s_i}^{(p)}$ with \textbf{counterfactual (CF) properties} $\mathbf P_{ij}^{(p)}$ from another candidate $G_{ij}$, $j\ne s_i$, in the same family, termed the \textbf{CF candidate}. Each eligible alternative defines a separate substitution, with prediction $\widehat s_{ij,\mathrm{CF}}^{(p)}$ evaluated against $j$. We assess whether a model switches to the corresponding CF candidate after property substitution. 

\subsection{Metrics}
\label{sec:metrics}
We use Property-Grounded Recovery as the primary metric, with Reliance Margin and Counterfactual Flip Rate as diagnostic metrics.

\subsubsection{Primary metric} \textbf{Property-Grounded Recovery (PGR)} jointly evaluates recovery under the original source property and candidate-targeted counterfactual substitutions, assigning equal weight to source and counterfactual recovery within each task family.

Let $\widehat{s}_{ij}^{(p)}$ denote the prediction when candidate $j$ supplies the property: the source prediction when $j=s_i$, and the CF prediction otherwise. With $n_i$ and $s_i$ defined in Section~\ref{sec:protocol},
\begin{equation}
\mathrm{PGR}_p=\frac{50}{|\mathcal E_p|}\sum_{i\in\mathcal E_p}
\left[
\mathbf{1}\!\left(\widehat{s}_{is_i}^{(p)}=s_i\right)
+\frac{1}{n_i-1}\sum_{j\ne s_i}\mathbf{1}\!\left(\widehat{s}_{ij}^{(p)}=j\right)
\right].
\label{eq:pgr}
\end{equation}
Here, $\mathcal E_p$ is the input-validated family set shared across all evaluated models for channel $p$, with source properties and CF properties available for every alternative candidate. Counterfactual correctness is first averaged within each family, ensuring that families with more candidates do not receive greater weight. Source recovery and family-level counterfactual recovery each contribute half of the score. Terminal-invalid outputs count as failures without reducing the denominator. PGR is reported separately for each channel as a percentage, without conditioning on source correctness or averaging across channels.

\subsubsection{Diagnostic metrics}

\noindent\textbf{Reliance Margin (RM)} measures the model's \textbf{reliance}: performance gain from providing the correctly paired property relative to a shuffled-property control. A positive RM indicates that the correct condition-target alignment explicitly improves selection accuracy (reported in percentage points in Table~\ref{tab:evidence-pairing-gain}).

\noindent\textbf{Counterfactual Flip Rate (CFR)} measures the model's \textbf{responsiveness}: how consistently a model flips to the corresponding CF candidate after property substitution, conditioned on a correct initial prediction. We compute the proportion of successful flips across eligible CF substitutions for these qualifying tasks, then average the results. CFR specifically quantifies responsiveness within each evaluated cohort. Table~\ref{tab:cf-transition} reports this metric as a percentage.

Formal definitions, eligibility criteria, and invalid-response handling for RM and CFR are provided in Appendix~\ref{app:metric-definitions}.

\subsection{Implementation}
\label{sec:implementation}
We align molecular records from QM9~\citep{ramakrishnan2014qm9}, QM9-NMR~\citep{gupta2021qm9nmr}, and QM9S~\citep{zou2023qm9s}. QM9 supplies the HOMO--LUMO gap, QM9-NMR supplies gas-phase NMR shieldings, and QM9S supplies dipole, IR, Raman, and UV--Vis observations.

We evaluate the 11 single-channel inputs defined in Section~\ref{sec:task}. IR and Raman use extracted peak lists, while UV--Vis supplies the first ten excitation states. Final cohort selection uses the property-specific checks in Appendix~\ref{app:construction-filter}. NMR filtering uses integration-augmented $^1$H and degeneracy-augmented $^{13}$C shieldings; the other NMR representations use the same selected families. Appendices~\ref{app:release} and~\ref{app:properties} document data alignment, observation scopes, units, and preprocessing. Appendix~\ref{app:property_background} explains the theoretical background of the properties.

Evaluation follows a closed protocol without tools, retrieval, task-specific training, or solved few-shot examples. Appendix~\ref{app:inference} provides inference settings, scoring procedures, and uncertainty details; Appendix~\ref{app:cf} documents CF coverage and eligibility.

\begin{table}[!b]
\centering
\caption{\textbf{Property-Grounded Recovery (PGR, \%) leaderboard.} Scores give equal weight to source and counterfactual recovery within each family (Eq.~\ref{eq:pgr}). Each channel uses the same complete-coverage, input-validated families across all seven models. Gold~\protect\goldcrown\ and silver~\protect\silvercrown\ mark the highest (bold) and second-highest model scores in each column. The random baseline is the expected score under uniform candidate selection.}
\label{tab:pgr-leaderboard}
\begingroup
\setlength{\tabcolsep}{3pt}
\setlength{\arrayrulewidth}{0.35pt}
\setlength{\doublerulesep}{1.5pt}
\renewcommand{\arraystretch}{1.15}
\fontsize{8}{9.6}\selectfont
\makeatletter
\newcommand{\pgrtablegroup}[2]{\smash{\raisebox{-#1\dimexpr\ht\@arstrutbox+\dp\@arstrutbox\relax}{$\vcenter{\hbox{#2}}$}}}
\makeatother
\resizebox{\linewidth}{!}{%
\begin{tabular}{@{}>{\hspace{4pt}}c<{\hspace{2pt}}|>{\hspace{2pt}}l<{\hspace{2pt}}|*{11}{c}@{}}
\hline\hline
\pgrtablegroup{1}{\textbf{Weights}} & \pgrtablegroup{1}{\textbf{Model}} & \multicolumn{3}{c}{\textbf{Electronic}} & \multicolumn{2}{c}{\textbf{Vibrational}} & \multicolumn{6}{c}{\textbf{Magnetic resonance}} \\
\cline{3-13}
 & & & & & & & \multicolumn{3}{c}{\textbf{$^1$H NMR}} & \multicolumn{3}{c}{\textbf{$^{13}$C NMR}} \\
\cline{8-13}
 & & \textbf{Dipole} & \textbf{Gap} & \textbf{UV--Vis} & \textbf{IR} & \textbf{Raman} & \textbf{Peaks} & \textbf{+Int.} & \textbf{Atomwise} & \textbf{Peaks} & \textbf{+Deg.} & \textbf{Atomwise} \\
\hline
\multicolumn{2}{c|}{Uniform random} & 37.9\nocrown & 37.7\nocrown & 37.9\nocrown & 37.9\nocrown & 37.9\nocrown & 37.7\nocrown & 37.7\nocrown & 37.7\nocrown & 37.7\nocrown & 37.7\nocrown & 37.7\nocrown \\
\hline
\pgrtablegroup{1}{Open} & Qwen3.8-27B & 44.2\nocrown & 35.9\nocrown & 38.6\nocrown & 58.2\nocrown & 49.5\nocrown & 35.5\nocrown & 36.0\nocrown & 38.0\nocrown & 37.7\nocrown & 37.0\nocrown & 39.5\nocrown \\
 & Gemma-4-31B-it & 45.5\nocrown & 37.9\nocrown & 46.0\nocrown & 68.6\nocrown & 55.0\nocrown & 37.8\nocrown & 36.1\nocrown & 37.8\nocrown & 36.9\nocrown & 36.9\nocrown & 38.6\nocrown \\
 & ChemDFM-v2.0-14B & 37.5\nocrown & 37.4\nocrown & 35.6\nocrown & 38.9\nocrown & 38.1\nocrown & 36.9\nocrown & 36.9\nocrown & 40.1\nocrown & 35.6\nocrown & 36.0\nocrown & 36.0\nocrown \\
\hline
\pgrtablegroup{1.5}{Closed} & GPT-5.6-Luna & 62.2\nocrown & 47.1\nocrown & 58.4\nocrown & 82.4\nocrown & 66.5\nocrown & 38.4\nocrown & 43.8\nocrown & 40.2\nocrown & 44.9\nocrown & 41.9\nocrown & 47.2\nocrown \\
 & GPT-5.6-Sol & 64.6\silvercrown & 69.0\silvercrown & 73.2\silvercrown & 90.7\silvercrown & 72.1\silvercrown & 55.2\nocrown & 57.5\silvercrown & 55.6\nocrown & 58.1\nocrown & 58.1\nocrown & 56.3\nocrown \\
 & Claude Opus 5 & 59.7\nocrown & 55.0\nocrown & 66.1\nocrown & 88.5\nocrown & 71.5\nocrown & 55.4\silvercrown & 55.0\nocrown & 58.8\silvercrown & 68.0\silvercrown & 62.5\silvercrown & \textbf{69.3}\goldcrown \\
 & GPT-6-Astra & \textbf{69.4}\goldcrown & \textbf{72.3}\goldcrown & \textbf{80.0}\goldcrown & \textbf{93.1}\goldcrown & \textbf{74.7}\goldcrown & \textbf{70.6}\goldcrown & \textbf{71.2}\goldcrown & \textbf{64.6}\goldcrown & \textbf{68.7}\goldcrown & \textbf{69.3}\goldcrown & 68.1\silvercrown \\
\hline\hline
\end{tabular}%
}
\endgroup
\end{table}

\section{Results and Analysis}
\label{sec:results}

\subsection{Evaluation Setup}
We evaluate GPT-5.6-Luna and GPT-5.6-Sol~\citep{openai2026gpt56release}, GPT-6-Astra~\citep{openai2026astrarelease}, Claude Opus 5~\citep{anthropic2026opus5release}, Qwen3.8-27B~\citep{qwen2026qwen38modelcard}, Gemma-4-31B-it~\citep{google2026gemma4modelcard}, and ChemDFM-v2.0-14B~\citep{opendfm2025chemdfmv2modelcard}, abbreviated Luna, Sol, Astra, Opus, Qwen, Gemma, and ChemDFM. The closed protocol uses no tools, retrieval, task-specific training, or solved few-shot examples. Only input-validated records enter scoring. Appendix~\ref{app:inference} provides model identifiers, inference settings, coverage, and output-handling details. Table~\ref{tab:pgr-leaderboard} reports PGR, while Tables~\ref{tab:evidence-pairing-gain} and~\ref{tab:cf-transition} report the RM and CFR diagnostics. All three metrics are reported separately by property channel and NMR representation.

\subsection{Main Results}
\label{sec:pgr-results}
\label{sec:diagnostic-results}
Table~\ref{tab:pgr-leaderboard} reports the primary PGR leaderboard for all seven models and eleven property channels. Each channel is evaluated on a common model-independent family set with complete source and CF coverage, without conditioning on initial source correctness.

\textbf{PGR provides a channel-wise leaderboard of source and counterfactual recovery.} Astra has the highest point estimate in ten of the eleven channels, while Opus leads on atomwise $^{13}$C NMR (69.3\% versus Astra's 68.1\%). IR yields the highest PGR for Luna, Sol, Opus, and Astra, at 82.4\%, 90.7\%, 88.5\%, and 93.1\%, respectively. Qwen and Gemma also achieve their highest scores with IR (58.2\% and 68.6\%), whereas their NMR scores remain close to the uniform-random baseline. ChemDFM scores range from 35.6\% to 40.1\% across channels. These results summarize source and CF recovery jointly; the following diagnostics examine dependence on correct property pairing and conditional responsiveness separately.

\begin{table}[!t]
\caption[RM and CFR on MolP2SBench]{\textbf{Reliance and responsiveness across seven models.} \textbf{(a) RM (pp)} is the accuracy difference between source and shuffled properties; \textbf{(b) CFR (\%)} is the task-averaged rate of flips to the CF candidate among initially correct sources. Gold~\protect\goldcrown\ and silver~\protect\silvercrown\ mark the highest (bold) and second-highest scores in each column, including ties. Metric definitions and evaluated cohorts appear in Appendix~\ref{app:cf}.}
\makeatletter
\begingroup
\protected@edef\@currentlabel{\thetable(a)}
\label{tab:evidence-pairing-gain}
\endgroup
\begingroup
\protected@edef\@currentlabel{\thetable(b)}
\label{tab:cf-transition}
\endgroup
\makeatother
\centering
\vspace{5pt}
\begingroup
\setlength{\tabcolsep}{3.0pt}
\setlength{\arrayrulewidth}{0.35pt}
\setlength{\doublerulesep}{1.5pt}
\renewcommand{\arraystretch}{1.28}
\fontsize{8}{9.6}\selectfont
\makeatletter
\newcommand{\tablegroup}[2]{\smash{\raisebox{-#1\dimexpr\ht\@arstrutbox+\dp\@arstrutbox\relax}{$\vcenter{\hbox{#2}}$}}}
\makeatother
\resizebox{\linewidth}{!}{%
\begin{tabular}{@{}>{\hspace{4pt}}c<{\hspace{2pt}}|>{\hspace{2pt}}c<{\hspace{2pt}}|>{\hspace{2pt}}l<{\hspace{2pt}}|*{11}{c}@{}}
\hline\hline
\tablegroup{1}{\textbf{Metric}} & \tablegroup{1}{\textbf{Weights}} & \tablegroup{1}{\textbf{Model}} & \multicolumn{3}{c}{\textbf{Electronic}} & \multicolumn{2}{c}{\textbf{Vibrational}} & \multicolumn{6}{c}{\textbf{Magnetic resonance}} \\
\cline{4-14}
 & & & & & & & & \multicolumn{3}{c}{\textbf{\textsuperscript{1}H NMR}} & \multicolumn{3}{c}{\textbf{\textsuperscript{13}C NMR}} \\
\cline{9-14}
 & & & \textbf{Dipole} & \textbf{Gap} & \textbf{UV--Vis} & \textbf{IR} & \textbf{Raman} & \textbf{Peaks} & \textbf{+Int.} & \textbf{Atomwise} & \textbf{Peaks} & \textbf{+Deg.} & \textbf{Atomwise} \\
\hline
\tablegroup{3}{\shortstack{\textbf{(a)}\\\textbf{RM}\\(pp)}} & \tablegroup{1}{Open} & Qwen3.8-27B & 6.7\nocrown & $-$6.7\nocrown & 3.7\nocrown & 15.4\nocrown & 15.4\nocrown & $-$4.4\nocrown & 4.4\nocrown & $-$3.0\nocrown & 4.4\nocrown & 0.0\nocrown & 8.1\nocrown \\
 &  & Gemma-4-31B-it & 11.9\nocrown & 2.2\nocrown & 4.4\nocrown & 27.9\nocrown & 19.9\nocrown & 0.0\nocrown & 0.0\nocrown & 6.7\nocrown & $-$0.7\nocrown & $-$2.2\nocrown & $-$1.5\nocrown \\
 &  & ChemDFM-v2.0-14B & 3.7\nocrown & $-$3.0\nocrown & $-$2.2\nocrown & 2.2\nocrown & 0.7\nocrown & $-$2.2\nocrown & 5.2\nocrown & 0.7\nocrown & 0.0\nocrown & $-$3.0\nocrown & $-$2.2\nocrown \\
\cline{2-14}
 & \tablegroup{1.5}{Closed} & GPT-5.6-Luna & 30.4\silvercrown & 8.1\nocrown & 14.1\nocrown & 39.7\nocrown & 30.1\nocrown & $-$0.7\nocrown & 9.6\nocrown & 17.0\nocrown & 7.4\nocrown & 13.3\nocrown & 15.6\nocrown \\
 &  & GPT-5.6-Sol & 28.1\nocrown & \multicolumn{1}{c}{\textbf{33.3}\goldcrown} & 32.6\silvercrown & 49.3\silvercrown & 37.5\silvercrown & 17.8\nocrown & 25.2\silvercrown & 25.2\nocrown & 12.6\nocrown & 17.8\nocrown & 10.4\nocrown \\
 &  & Claude Opus 5 & 23.7\nocrown & 16.3\nocrown & 28.9\nocrown & 47.1\nocrown & 36.8\nocrown & 25.2\silvercrown & 13.3\nocrown & 28.9\silvercrown & \multicolumn{1}{c}{\textbf{28.9}\goldcrown} & \multicolumn{1}{c}{\textbf{27.4}\goldcrown} & \multicolumn{1}{c}{\textbf{37.0}\goldcrown} \\
 &  & GPT-6-Astra & \multicolumn{1}{c}{\textbf{36.3}\goldcrown} & 31.1\silvercrown & \multicolumn{1}{c}{\textbf{41.5}\goldcrown} & \multicolumn{1}{c}{\textbf{52.2}\goldcrown} & \multicolumn{1}{c}{\textbf{44.9}\goldcrown} & \multicolumn{1}{c}{\textbf{36.3}\goldcrown} & \multicolumn{1}{c}{\textbf{35.6}\goldcrown} & \multicolumn{1}{c}{\textbf{34.8}\goldcrown} & 22.2\silvercrown & 23.0\silvercrown & 18.5\silvercrown \\
\hline\hline
\tablegroup{3}{\shortstack{\textbf{(b)}\\\textbf{CFR}\\(\%)}} & \tablegroup{1}{Open} & Qwen3.8-27B & 26.4\nocrown & 11.0\nocrown & 16.7\nocrown & 46.3\nocrown & 33.3\nocrown & 11.6\nocrown & 7.1\nocrown & 14.6\nocrown & 10.3\nocrown & 7.3\nocrown & 12.3\nocrown \\
 &  & Gemma-4-31B-it & 31.7\nocrown & 6.0\nocrown & 28.7\nocrown & 61.4\nocrown & 34.2\nocrown & 6.2\nocrown & 5.4\nocrown & 6.4\nocrown & 5.8\nocrown & 11.2\nocrown & 5.0\nocrown \\
 &  & ChemDFM-v2.0-14B & 13.3\nocrown & 7.8\nocrown & 10.9\nocrown & 10.5\nocrown & 10.9\nocrown & 12.6\nocrown & 9.4\nocrown & 14.8\nocrown & 6.0\nocrown & 5.1\nocrown & 4.8\nocrown \\
\cline{2-14}
 & \tablegroup{1.5}{Closed} & GPT-5.6-Luna & 66.5\nocrown & 48.2\nocrown & 60.0\nocrown & 80.3\nocrown & 56.2\nocrown & 36.3\nocrown & 32.8\nocrown & 31.9\nocrown & 39.1\nocrown & 26.9\nocrown & 37.2\nocrown \\
 &  & GPT-5.6-Sol & 73.2\silvercrown & 63.6\silvercrown & 72.2\silvercrown & 90.1\silvercrown & 63.8\silvercrown & 54.7\nocrown & 55.4\silvercrown & 53.6\silvercrown & 67.3\silvercrown & 61.1\nocrown & 56.8\nocrown \\
 &  & Claude Opus 5 & 64.8\nocrown & 49.3\nocrown & 61.6\nocrown & 89.8\nocrown & 58.7\nocrown & 56.2\silvercrown & 50.9\nocrown & 52.9\nocrown & 66.5\nocrown & 65.2\silvercrown & 68.8\silvercrown \\
 &  & GPT-6-Astra & \multicolumn{1}{c}{\textbf{74.5}\goldcrown} & \multicolumn{1}{c}{\textbf{76.6}\goldcrown} & \multicolumn{1}{c}{\textbf{81.1}\goldcrown} & \multicolumn{1}{c}{\textbf{91.2}\goldcrown} & \multicolumn{1}{c}{\textbf{65.1}\goldcrown} & \multicolumn{1}{c}{\textbf{76.7}\goldcrown} & \multicolumn{1}{c}{\textbf{77.5}\goldcrown} & \multicolumn{1}{c}{\textbf{67.4}\goldcrown} & \multicolumn{1}{c}{\textbf{70.7}\goldcrown} & \multicolumn{1}{c}{\textbf{70.1}\goldcrown} & \multicolumn{1}{c}{\textbf{69.0}\goldcrown} \\
\hline\hline
\end{tabular}%
}
\endgroup
\end{table}

We distinguish two aspects of property use: whether correct structure--property pairing improves source recovery (Table~\ref{tab:evidence-pairing-gain}), and whether replacing those properties redirects an initially correct selection toward the counterfactual candidate (Table~\ref{tab:cf-transition}). Both interventions hold each task's scaffold and structural alternatives fixed.

\textbf{Models use physical properties, but performance varies across property channels.} All seven models show positive IR RM values, with IR yielding the largest or joint-largest gain for six of them. Within the GPT family, Luna, Sol, and Astra achieve IR gains of 39.7, 49.3, and 52.2 pp, respectively; Opus reaches 47.1 pp, close to Sol, and has the largest gains across the three $^{13}$C representations. Under IR substitution, CFR values are 80.3\% for Luna and 89.8--91.2\% for Sol, Opus, and Astra, evaluated on each model's initially correct sources with valid CF responses. Among open-weight models, Qwen and Gemma also benefit most from vibrational properties, with IR RM values of 15.4 and 27.9 pp, while several NMR views yield near-zero or negative gains. The chemistry-specialized ChemDFM shows more limited effects: its IR RM is 2.2 pp and its CFR is 10.5\%. Thus, the results indicate property-dependent selection across model groups, but its strength varies substantially by model and channel.

\textbf{Reliance and responsiveness show different observed property rankings.} Recall that RM quantifies a model's static reliance on correct property-structure pairing, whereas CFR measures its dynamic responsiveness to counterfactual evidence. Luna, Sol, and Astra all show a higher RM for atomwise $^1$H than for $^{13}$C peaks, but a lower CFR. For Sol, the RM values are 25.2 versus 12.6 pp, whereas the CFR values are 53.6\% versus 67.3\%. CFR is evaluated on each view's initially correct source cohort. This observed reversal motivates reporting reliance and responsiveness separately: RM measures the benefit of correct property pairing, whereas CFR measures responses to targeted property substitutions. Disentangling these dimensions may inform downstream applications, such as human-in-the-loop molecular design, where models must adapt to updated experimental feedback. 

\textbf{LLMs can perform better with properties that are less intuitive to chemists.} NMR offers familiar links between spectral features and local chemical environments, whereas UV--Vis provides less direct structural information. Nevertheless, Astra achieves 80.1\% source recovery with UV--Vis, compared with 69.9\% and 66.2\% with $^1$H and $^{13}$C NMR peaks, respectively. Sol shows the same ordering. These observed results highlight UV--Vis as a useful input for molecular selection alongside the more familiar NMR representations.

In summary, these results establish property-dependent selection, while leaving the validity of the supporting chemical argument open. After examining combined-property inputs and the interpretation of model comparisons, we use case studies to examine how models connect the supplied observations to candidate-specific environments and geometry.

\begin{figure}[!t]
\centering
\includegraphics[width=\linewidth]{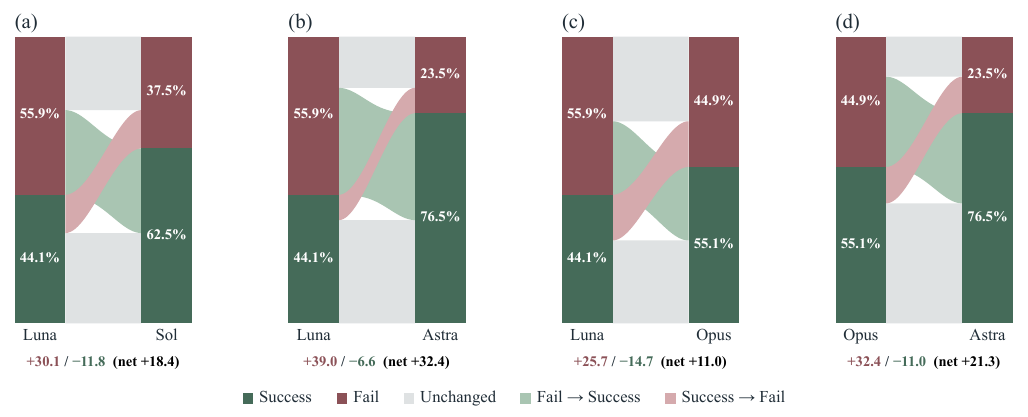}
\caption{Success/failure transitions across four pairs of closed-weight models under NMR-bundle input: (a) Luna--Sol, (b) Luna--Astra, (c) Luna--Opus, and (d) Opus--Astra. Percentages refer to all tasks; $+/-$ indicate gains/losses from left to right, and net reports the difference in source-property accuracy in pp.}
\label{fig:model-bundle-nmr}
\end{figure}

\subsection{Combined Property Analysis}
\label{sec:combined-property}
We compare single properties, related-property bundles, and all-properties input while keeping constituent values and candidate order matched (Appendix~\ref{app:combined}).

\phantomsection\label{sec:within-bundle}
\textbf{Combining properties does not always improve recovery.} Within the GPT family, Sol and Astra improve over the strongest constituent in all three bundles, with larger gains for Astra; for example, electronic recovery rises from 72.8\% to 73.5\% for Sol and from 80.1\% to 88.2\% for Astra. Luna instead improves only with IR and Raman (81.6\% to 83.8\%), while electronic and NMR recovery fall from 65.4\% to 61.8\% and from 49.3\% to 44.1\%, respectively. Opus also benefits from electronic and vibrational combinations, but its NMR recovery drops from 66.2\% to 55.1\%. Thus, combining related channels can help, but does not consistently outperform the strongest individual input.

\phantomsection\label{sec:across-bundle}
\textbf{A successful within-bundle combination does not guarantee a successful cross-bundle combination.} Among closed-weight models, adding the remaining properties to the strongest bundle (vibrational for all four) improves Sol from 89.7\% to 93.4\% and Astra from 96.3\% to 98.5\%, but reduces Luna from 83.8\% to 78.7\% and Opus from 90.4\% to 89.0\%. Thus, Luna and Opus benefit from combining IR and Raman but not from extending that bundle to all properties. Among open-weight models, Qwen and Gemma show the same reversal: both benefit from IR and Raman together, yet recovery falls from 64.0\% to 48.5\% and from 74.3\% to 54.4\%, respectively, when the remaining families are added. The single-property--bundle--all-properties progression is therefore not generally monotonic; more comprehensive input need not produce better recovery.

\begin{figure}[!t]
\centering
\begingroup
\definecolor{casecorrect}{HTML}{267D50}
\definecolor{casewrong}{HTML}{BE3030}
\sbox0{%
  \small
  \setlength{\tabcolsep}{6pt}%
  \renewcommand{\arraystretch}{1.12}%
  \begin{tabular}{@{}lcc@{}}
  \toprule
  \textbf{Model} & \textbf{Given ester $^{13}$C} & \textbf{Given ether $^{13}$C} \\
  \midrule
  \textbf{Astra} & \textcolor{casecorrect}{Ester $\checkmark$} & \textcolor{casecorrect}{Ether $\checkmark$} \\
  Luna & \textcolor{casewrong}{Ether $\times$} & \textcolor{casewrong}{Ester $\times$} \\
  Sol & \textcolor{casewrong}{Ether $\times$} & \textcolor{casecorrect}{Ether $\checkmark$} \\
  Opus & \textcolor{casewrong}{Ether $\times$} & \textcolor{casecorrect}{Ether $\checkmark$} \\
  \bottomrule
  \end{tabular}%
}
\begin{minipage}[t]{0.46\linewidth}
\vspace{0pt}\centering
\includegraphics[width=\linewidth,height=\dimexpr\ht0+\dp0\relax,keepaspectratio]{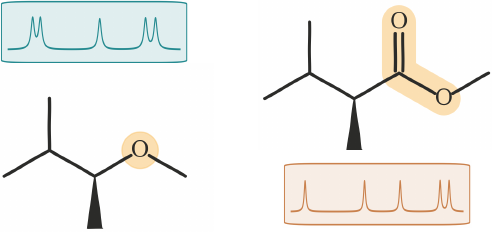}
\caption{Ether (left) and ester~(right). Orange highlights the candidate fragments.}
\label{fig:case-nmr-structures}
\end{minipage}\hfill
\begin{minipage}[t]{0.50\linewidth}
\vspace{0pt}\centering
\makeatletter\def\@captype{table}\makeatother
\setlength{\abovecaptionskip}{0pt}\caption{Predictions under each $^{13}$C input. Checkmarks indicate selection of the property-providing candidate.}
\label{tab:case-nmr-predictions}
\usebox0
\end{minipage}
\endgroup
\end{figure}

\subsection{Do Recovery-Rate Differences Reflect Capability Differences across Models?}
\label{sec:recovery-capability}

\textbf{Higher recovery does not subsume another model's successes.} The paired NMR-bundle results in Figure~\ref{fig:model-bundle-nmr} show that improvements in overall recovery accompany changes in both directions. Sol achieves 62.5\% recovery compared with Luna's 44.1\%, but 30.1\% of all tasks are solved by Sol and not Luna, while 11.8\% are solved by Luna and not Sol. The contrast is sharper for Opus: it improves recovery to 55.1\%, yet 25.7\% of tasks are solved by Opus and not Luna, and 14.7\% by Luna and not Opus. Across the four paired NMR comparisons, net gains range from 11.0 to 32.4 pp, whereas the churn rate, defined as the share of all tasks whose success/failure outcome differs between the two models, ranges from 40.4\% to 45.6\%, so even a large recovery advantage does not amount to a simple expansion of the comparator's solved task set.

\textbf{Bundle-level advantages do not necessarily extend to single properties.} Opus exceeds Astra on $^{13}$C peaks-only source recovery (69.1\% versus 66.2\%) and atomwise source recovery (70.6\% versus 66.9\%), despite its lower NMR-bundle recovery (55.1\% versus 76.5\%). This source-recovery advantage does not carry over to unconditional CF accuracy: Astra exceeds Opus with both peaks-only input (65.9\% versus 63.6\%) and atomwise input (65.1\% versus 64.0\%; Appendix~\ref{app:cf}, Table~\ref{tab:counterfactual-recovery}). Model rankings therefore depend on both the supplied properties and the diagnostic being evaluated. 

\subsection{Case Study}
\label{sec:case-study}

\label{sec:nmr-scope}
\label{sec:geometry-cases}
\textbf{What Astra's Advantage Reveals about Property Interpretation and Molecular Structure Understanding.} Task \hyperref[app:family-97ccf0b132ed0cc7]{97CCF0B132ED0CC7} (\hyperref[app:task-nmr-97ccf]{Appendix~\ref*{app:nmr-details}}) contrasts the ether \path{CO[C@@H](C)C(C)C} with the ester \path{COC(=O)[C@@H](C)C(C)C}, whose property values are supplied. The retained methine carbon is directly oxygen-bound in the ether but adjacent to the carbonyl carbon in the ester. Given $^{13}$C shieldings of 134.3052--168.7333~ppm, Astra selects the ester, arguing that the ether's oxygen-bound methine should be more deshielded and explicitly noting that the masked carbonyl carbon is not observed. Luna and Opus select the ether, invoking the absence or expected contribution of a carbonyl signal; Sol also selects the ether, assigning the two lowest shieldings to two oxygen-bound retained carbons. The distinction therefore requires connecting candidate-specific connectivity to the actual observation scope, not merely recognizing that an ester contains a carbonyl.

\noindent When the property input is replaced by the ether's scaffold shieldings, including 101.2638 and 130.2122~ppm, Astra switches to the ether and attributes the stronger deshielding to its oxygen-bound methine and methoxy carbons. Only Astra selects the property-providing candidate in both conditions: Luna reverses the two assignments, while Sol and Opus select the ether throughout.

\textbf{Models make more errors when property values conflict with local heuristics.} For task \hyperref[app:family-49e99a99315e5e96]{49E99A99315E5E96} (\hyperref[app:task-dipole-49e99]{Appendix~\ref*{app:geometry-diagnostics}}), the carbonyl-containing target has a dipole of 1.45724~D, below the reduced candidate's 5.37590~D. All four models select the reduced candidate under dipole-magnitude input; Astra argues that the additional carbonyl is less consistent with the low magnitude. The reference values contradict this local polarity heuristic. Across 35 within-task matched carbonyl/reduced pairs, ten similarly reverse the expectation that adding a carbonyl increases molecular dipole. Astra makes errors on 7/10 reversal tasks versus 5/25 nonreversal tasks; Luna, Sol, and Opus also have higher error rates in the reversal subset (Appendix~\ref{app:geometry-diagnostics}). These post-hoc contrasts identify a limitation of the heuristic, not geometry as its isolated cause: the substitutions change electronic structure and conformation together.

\section{Conclusion and Limitation}
\label{sec:discussion}

MolP2SBench establishes a controlled framework for evaluating property-to-structure reasoning in LLMs beyond answer accuracy. Across seven models, reliance and responsiveness rank property channels differently, additional properties can reduce recovery, and case studies reveal physical inconsistencies in model explanations. By combining curated completion tasks, a channel-wise recovery leaderboard, and matched property interventions, MolP2SBench provides the community with a reusable testbed for diagnosing evidence use and evaluating future training strategies toward more reliable property-grounded molecular inference.

\textbf{Limitation.} MolP2SBench covers small-molecule completion with computed properties and curated candidates; experimental spectra and open-ended structure elucidation remain outside its scope. The current benchmark’s coverage is constrained by the effort required for careful dataset construction and manual validation. Future work could expand molecular families and property channels through LLM-assisted dataset construction while retaining expert review and validation.

\bibliography{references}

\begin{thebibliography}{34}
\providecommand{\natexlab}[1]{#1}
\providecommand{\url}[1]{\texttt{#1}}
\expandafter\ifx\csname urlstyle\endcsname\relax
  \providecommand{\doi}[1]{doi: #1}\else
  \providecommand{\doi}{doi: \begingroup \urlstyle{rm}\Url}\fi

\bibitem[{Anthropic}(2026)]{anthropic2026opus5release}
{Anthropic}.
\newblock Introducing {Claude Opus 5}, July 2026.
\newblock URL \url{https://www.anthropic.com/news/claude-opus-5}.
\newblock Official release announcement.

\bibitem[Baerends et~al.(2013)Baerends, Gritsenko, and Van~Meer]{baerends2013gaps}
EJ~Baerends, OV~Gritsenko, and R~Van~Meer.
\newblock The kohn--sham gap, the fundamental gap and the optical gap: the physical meaning of occupied and virtual kohn--sham orbital energies.
\newblock \emph{Physical Chemistry Chemical Physics}, 15\penalty0 (39):\penalty0 16408--16425, 2013.

\bibitem[Bartmann et~al.(2026)Bartmann, Schimunek, Ielanskyi, Seidl, Klambauer, and Luukkonen]{bartmann2026moleculariq}
Christoph Bartmann, Johannes Schimunek, Mykyta Ielanskyi, Philipp Seidl, G{\"u}nter Klambauer, and Sohvi Luukkonen.
\newblock Moleculariq: Characterizing chemical reasoning capabilities through symbolic verification on molecular graphs.
\newblock 2026.

\bibitem[Born \& Manica(2023)Born and Manica]{born2023regression}
Jannis Born and Matteo Manica.
\newblock Regression transformer enables concurrent sequence regression and generation for molecular language modelling.
\newblock \emph{Nature Machine Intelligence}, 5\penalty0 (4):\penalty0 432--444, 2023.

\bibitem[Bushuiev et~al.(2024)Bushuiev, Bushuiev, de~Jonge, Young, Kretschmer, Samusevich, Heirman, Wang, Zhang, D{\"u}hrkop, et~al.]{bushuiev2024massspecgym}
Roman Bushuiev, Anton Bushuiev, Niek~F de~Jonge, Adamo Young, Fleming Kretschmer, Raman Samusevich, Janne Heirman, Fei Wang, Luke Zhang, Kai D{\"u}hrkop, et~al.
\newblock Massspecgym: A benchmark for the discovery and identification of molecules.
\newblock volume~37, pp.\  110010--110027, 2024.

\bibitem[Cai et~al.(2026)Cai, Bai, Tang, He, Luo, Zhu, Pilla, Li, Liu, and Luo]{cai2026mollangbench}
Feiyang Cai, Jiahui Bai, Tao Tang, Guijuan He, Joshua Luo, Tianyu Zhu, Srikanth Pilla, Gang Li, Ling Liu, and Feng Luo.
\newblock Mollangbench: A comprehensive benchmark for language-prompted molecular structure recognition, editing, and generation.
\newblock In \emph{International Conference on Learning Representations}, volume 2026, pp.\  83272--83309, 2026.

\bibitem[Feng et~al.(2019)Feng, Wallace, and Boyd-Graber]{feng2019misleading}
Shi Feng, Eric Wallace, and Jordan~Lee Boyd-Graber.
\newblock Misleading failures of partial-input baselines.
\newblock In \emph{Proceedings of the 57th Annual Meeting of the Association for Computational Linguistics}, pp.\  5533--5538, 2019.

\bibitem[Gardner et~al.(2020)Gardner, Artzi, Basmov, Berant, Bogin, Chen, Dasigi, Dua, Elazar, Gottumukkala, et~al.]{gardner2020contrast}
Matt Gardner, Yoav Artzi, Victoria Basmov, Jonathan Berant, Ben Bogin, Sihao Chen, Pradeep Dasigi, Dheeru Dua, Yanai Elazar, Ananth Gottumukkala, et~al.
\newblock Evaluating models’ local decision boundaries via contrast sets.
\newblock In \emph{Findings of the Association for Computational Linguistics: EMNLP 2020}, pp.\  1307--1323, 2020.

\bibitem[Geirhos et~al.(2020)Geirhos, Jacobsen, Michaelis, Zemel, Brendel, Bethge, and Wichmann]{geirhos2020shortcut}
Robert Geirhos, J{\"o}rn-Henrik Jacobsen, Claudio Michaelis, Richard Zemel, Wieland Brendel, Matthias Bethge, and Felix~A Wichmann.
\newblock Shortcut learning in deep neural networks.
\newblock \emph{Nature Machine Intelligence}, 2\penalty0 (11):\penalty0 665--673, 2020.

\bibitem[{Google DeepMind}(2026)]{google2026gemma4modelcard}
{Google DeepMind}.
\newblock {Gemma-4-31B-it}.
\newblock Hugging Face model card, 2026.
\newblock URL \url{https://huggingface.co/google/gemma-4-31B-it}.

\bibitem[Guo et~al.(2026)Guo, Li, Cao, Zhang, and Yuan]{guo2026chemcotv2}
Hongyu Guo, Hao Li, He~Cao, Gongbo Zhang, and Li~Yuan.
\newblock From answers to states: Verifiable process-level evaluation of chemical reasoning in large language models.
\newblock \emph{arXiv preprint arXiv:2606.03660}, 2026.

\bibitem[Guo et~al.(2024)Guo, Nan, Zhou, Guo, Guo, Surve, Liang, Chawla, Wiest, and Zhang]{guo2024molpuzzle}
Kehan Guo, Bozhao Nan, Yujun Zhou, Taicheng Guo, Zhichun Guo, Mihir Surve, Zhenwen Liang, Nitesh~V Chawla, Olaf Wiest, and Xiangliang Zhang.
\newblock Can llms solve molecule puzzles? a multimodal benchmark for molecular structure elucidation.
\newblock volume~37, pp.\  134721--134746, 2024.

\bibitem[Gupta et~al.(2021)Gupta, Chakraborty, and Ramakrishnan]{gupta2021qm9nmr}
Amit Gupta, Sabyasachi Chakraborty, and Raghunathan Ramakrishnan.
\newblock Revving up 13c nmr shielding predictions across chemical space: benchmarks for atoms-in-molecules kernel machine learning with new data for 134 kilo molecules.
\newblock \emph{Machine Learning: Science and Technology}, 2\penalty0 (3):\penalty0 035010, 2021.

\bibitem[Han et~al.(2026)Han, Wu, Wang, Zhou, Lv, Zhao, and Hu]{han2026molquest}
Taolin Han, Shuang Wu, Jinghang Wang, Yuhao Zhou, Renquan Lv, Bing Zhao, and Wei Hu.
\newblock Molquest: A benchmark for agentic evaluation of abductive reasoning in chemical structure elucidation.
\newblock \emph{arXiv preprint arXiv:2603.25253}, 2026.

\bibitem[Hao et~al.(2026)Hao, Cao, Feng, Shao, Tang, Yan, Tian, Yuan, and Li]{li2025chemcot}
Li~Hao, He~Cao, Bin Feng, Daniel Shao, Robert Tang, Zhiyuan Yan, Yonghong Tian, Li~Yuan, and Yu~Li.
\newblock Beyond chemical qa: Evaluating llm's chemical reasoning with modular chemical operations.
\newblock volume~38, 2026.

\bibitem[Inae et~al.(2026)Inae, Gunn, Bond, and Jiang]{inae2026chemdirt}
Eric Inae, Tim Gunn, Chris Bond, and Meng Jiang.
\newblock Chemdirt: A diversified instruction, representation, and task benchmark for robust chemistry-llm evaluation.
\newblock \emph{arXiv preprint arXiv:2608.21504}, 2026.

\bibitem[Jinger et~al.(2021)Jinger, Fliegl, Bast, Dimitrova, Lehtola, and Sundholm]{jinger2021shieldings}
Rahul~Kumar Jinger, Heike Fliegl, Radovan Bast, Maria Dimitrova, Susi Lehtola, and Dage Sundholm.
\newblock Spatial contributions to nuclear magnetic shieldings.
\newblock \emph{The Journal of Physical chemistry. A}, 125\penalty0 (8):\penalty0 1778, 2021.

\bibitem[Liu et~al.(2026)Liu, Ouyang, Zhong, Han, and Zhao]{liu2025fgbench}
Xuan Liu, Siru Ouyang, Xianrui Zhong, Jiawei Han, and Huimin Zhao.
\newblock Fgbench: A dataset and benchmark for molecular property reasoning at functional group-level in large language models.
\newblock volume~38, 2026.

\bibitem[Mahmood et~al.(2021)Mahmood, Mansimov, Bonneau, and Cho]{mahmood2021masked}
Omar Mahmood, Elman Mansimov, Richard Bonneau, and Kyunghyun Cho.
\newblock Masked graph modeling for molecule generation.
\newblock \emph{Nature communications}, 12\penalty0 (1):\penalty0 3156, 2021.

\bibitem[Noh et~al.(2026)Noh, Lee, Na, Kim, and Park]{noh2026iragent}
Heewoong Noh, Namkyeong Lee, Gyoung~S Na, Kibum Kim, and Chanyoung Park.
\newblock Ir-agent: Expert-inspired llm agents for structure elucidation from infrared spectra.
\newblock In \emph{International Conference on Learning Representations}, volume 2026, pp.\  14555--14583, 2026.

\bibitem[{OpenAI}(2026{\natexlab{a}})]{openai2026astrarelease}
{OpenAI}.
\newblock {GPT-6 Astra}: A new generation of intelligence, September 2026{\natexlab{a}}.
\newblock URL \url{https://openai.com/index/gpt-6-astra/}.
\newblock Official release announcement.

\bibitem[{OpenAI}(2026{\natexlab{b}})]{openai2026gpt56release}
{OpenAI}.
\newblock {GPT-5.6}: Frontier intelligence that scales with your ambition, July 2026{\natexlab{b}}.
\newblock URL \url{https://openai.com/index/gpt-5-6/}.
\newblock Official release announcement.

\bibitem[{OpenDFM}(2025)]{opendfm2025chemdfmv2modelcard}
{OpenDFM}.
\newblock {ChemDFM-v2.0-14B}.
\newblock Hugging Face model card, 2025.
\newblock URL \url{https://huggingface.co/OpenDFM/ChemDFM-v2.0-14B}.

\bibitem[{Qwen Team}(2026)]{qwen2026qwen38modelcard}
{Qwen Team}.
\newblock {Qwen3.8-27B}.
\newblock Hugging Face model card, 2026.
\newblock URL \url{https://huggingface.co/Qwen/Qwen3.8-27B}.

\bibitem[Ramakrishnan et~al.(2014)Ramakrishnan, Dral, Rupp, and Von~Lilienfeld]{ramakrishnan2014qm9}
Raghunathan Ramakrishnan, Pavlo~O Dral, Matthias Rupp, and O~Anatole Von~Lilienfeld.
\newblock Quantum chemistry structures and properties of 134 kilo molecules.
\newblock \emph{Scientific data}, 1\penalty0 (1):\penalty0 140022, 2014.

\bibitem[Rothchild et~al.(2021)Rothchild, Tamkin, Yu, Misra, and Gonzalez]{rothchild2021c5t5}
Daniel Rothchild, Alex Tamkin, Julie Yu, Ujval Misra, and Joseph Gonzalez.
\newblock C5t5: Controllable generation of organic molecules with transformers.
\newblock \emph{arXiv preprint arXiv:2108.10307}, 2021.

\bibitem[Runcie et~al.(2026)Runcie, Deane, and Imrie]{runcie2026chemiq}
Nicholas~T Runcie, Charlotte~M Deane, and Fergus Imrie.
\newblock Assessing the chemical intelligence of large language models.
\newblock \emph{Journal of chemical information and modeling}, 66\penalty0 (1):\penalty0 216--227, 2026.

\bibitem[Su et~al.(2025)Su, Chen, Jiang, Zhong, Wang, Liu, and Zhang]{su2026spectrallm}
Yunyue Su, Jiahui Chen, Zao Jiang, Zhenyi Zhong, Liang Wang, Qiang Liu, and Zhaoxiang Zhang.
\newblock Spectrallm: Uncovering the ability of llms for molecular structure elucidation from multi-spectral data.
\newblock 2025.

\bibitem[Wilson et~al.(1955)Wilson, Decius, Cross, et~al.]{wilson1980vibrations}
Edgar~Bright Wilson, John~Courtney Decius, Paul~C Cross, et~al.
\newblock \emph{Molecular vibrations: the theory of infrared and Raman vibrational spectra}, volume 102.
\newblock McGraw-Hill New York, 1955.

\bibitem[Xiang et~al.(2026)Xiang, Ma, Chen, Wang, Chen, and Zeng]{xiang2026specx}
Chengrui Xiang, Tengfei Ma, Yujie Chen, Tong Wang, Haowen Chen, and Xiangxiang Zeng.
\newblock Specx: A large-scale benchmark for multi-modal spectroscopy and cross-paradigm evaluation.
\newblock \emph{arXiv preprint arXiv:2605.18791}, 2026.

\bibitem[Yan et~al.(2025)Yan, Chen, and Cho]{yan2025inconsistency}
Bing Yan, Angelica Chen, and Kyunghyun Cho.
\newblock Inconsistency of llms in molecular representations.
\newblock \emph{Digital Discovery}, 4\penalty0 (10):\penalty0 2876--2892, 2025.

\bibitem[Yang et~al.(2026)Yang, Xie, Shen, Wang, Chen, Gao, Sun, Qi, Zhou, Bai, et~al.]{yang2026spectrumworld}
Zhuo Yang, Jiaqing Xie, Shuaike Shen, Daolang Wang, Yeyun Chen, Ben Gao, Shuzhou Sun, Biqing Qi, Dongzhan Zhou, Lei Bai, et~al.
\newblock Spectrumworld: Artificial intelligence foundation for spectroscopy.
\newblock In \emph{Proceedings of the 32nd ACM SIGKDD Conference on Knowledge Discovery and Data Mining V. 2}, pp.\  12655--12666, 2026.

\bibitem[Ye et~al.(2025)Ye, Cai, Lai, Wang, Huang, Wang, Liu, and Zeng]{ye2025drugassist}
Geyan Ye, Xibao Cai, Houtim Lai, Xing Wang, Junhong Huang, Longyue Wang, Wei Liu, and Xiangxiang Zeng.
\newblock Drugassist: A large language model for molecule optimization.
\newblock \emph{Briefings in Bioinformatics}, 26\penalty0 (1):\penalty0 bbae693, 2025.

\bibitem[Zou et~al.(2023)Zou, Zhang, Liang, Wei, Leng, Jiang, Luo, and Hu]{zou2023qm9s}
Zihan Zou, Yujin Zhang, Lijun Liang, Mingzhi Wei, Jiancai Leng, Jun Jiang, Yi~Luo, and Wei Hu.
\newblock A deep learning model for predicting selected organic molecular spectra.
\newblock \emph{Nature Computational Science}, 3\penalty0 (11):\penalty0 957--964, 2023.

\end{thebibliography}
\bibliographystyle{iclr2027_conference}
\clearpage
\appendix
\startcontents[appendices]

\section*{Supplementary Material}
Appendix~\ref{sec:related} provides the extended related-work discussion. Appendices~\ref{app:release}--\ref{app:inference} document construction, physical background, representations, and the current execution snapshot. Appendices~\ref{app:repr}--\ref{app:combined} give source-recovery, conditional-transition, and paired-combination results. Appendix~\ref{app:cases} documents verified case records and exploratory geometry diagnostics. Appendix~\ref{app:nmr-details} gives NMR cohort definitions, paired statistics, and selected cases, and Appendix~\ref{app:task-families} lists all final-cohort source tasks, candidate structures, and scaffold drawings.

\subsection*{Appendix Contents}
\begingroup
\small
\setlength{\parskip}{0pt}
\printcontents[appendices]{}{1}{\setcounter{tocdepth}{2}}
\endgroup
\clearpage

\section{Related Work}
\label{sec:related}

Table~\ref{tab:related_work_comparison} compares the benchmarks discussed below.

\paragraph{Molecular reasoning and structure elucidation.}
FGBench evaluates functional-group property reasoning~\citep{liu2025fgbench}; ChemCoTBench and MolLangBench test molecular operations~\citep{li2025chemcot,cai2026mollangbench}; ChemIQ examines structural reasoning and NMR elucidation~\citep{runcie2026chemiq}. Spectrum-to-structure studies address generation, retrieval, and interactive property acquisition~\citep{guo2024molpuzzle,su2026spectrallm,bushuiev2024massspecgym,han2026molquest,xiang2026specx,yang2026spectrumworld}. MolP2SBench fixes a masked scaffold and complete candidate molecules, then substitutes aligned physical observations to test whether decisions follow the evidence.

\paragraph{Property-conditioned reconstruction.}
Masked graph modeling reconstructs molecular graphs under property conditions~\citep{mahmood2021masked}; C5T5 reconstructs corrupted molecular names conditioned on properties~\citep{rothchild2021c5t5}; and the Regression Transformer combines property prediction with conditioned generation~\citep{born2023regression}. These approaches develop models through conditional reconstruction. We instead use local completion to evaluate general-purpose language models, constructing candidate-aligned observations that enable controlled property substitutions. Our contribution lies in the benchmark formulation and matched evaluation design, rather than masking itself.

\paragraph{Diagnostic evaluation and verification.}
MolecularIQ uses graph-derived symbolic verification~\citep{bartmann2026moleculariq}; consistency studies and ChemDIRT vary molecular representations or equivalent instructions~\citep{yan2025inconsistency,inae2026chemdirt}; and ChemCoTBench V2 evaluates intermediate chemical states~\citep{guo2026chemcotv2}. MolP2SBench changes the supplied property values while preserving the structural hypotheses: an appropriate response may therefore require changing the answer. We connect these behavioral comparisons to candidate-level checks of physical compatibility and stated explanations, separating correct selection from the validity of the argument supporting it.

\section{Dataset Construction}
\label{app:release}
\subsection{Candidate Families and Selection}
The final cohort contains 136 source tasks and 394 distinct candidate SMILES, with 58 two-candidate, 34 three-candidate, and 44 four-candidate tasks. There are 62 single-attachment and 74 two-attachment masks. Hidden connected groups contain one to three heavy atoms. Candidates are validated local completions, not structures generated by the evaluated models.

We group molecules by their canonical masked scaffold and attachment constraints, assemble structurally distinct completions, and select molecule-disjoint families spanning different fragment sizes and structural contrasts. We then align candidate properties and apply the property-specific checks below to determine the final cohort.

\subsection{Property-Specific Filtering}
\label{app:construction-filter}
For property channel $p$, let $\mathcal V_{ip}$ contain the unordered candidate pairs in family $i$ for which both aligned property records are available. A family is retained when its source-property inputs are complete and every verifiable pair passes the corresponding channel threshold:
\begin{equation}
 d_p\!\left(\mathbf P_{ij}^{(p)},\mathbf P_{ik}^{(p)}\right)\geq\tau_p
 \quad\text{for every }p\text{ and }(j,k)\in\mathcal V_{ip}.
\end{equation}
Thus, a verified below-threshold pair excludes the family.

\begin{table}[htbp]
\caption{Property-specific criteria for final cohort selection. Each threshold applies to every verifiable within-family candidate pair.}
\label{tab:construction-thresholds}
\centering
\begin{tabular}{lll}
\toprule
Property & Distance & Minimum \\
\midrule
QM9S dipole norm & Absolute magnitude difference & 0.30 D \\
HOMO--LUMO gap & Absolute difference & 0.0005 Hartree \\
$^1$H NMR & Shielding-distribution Wasserstein distance & 0.02 ppm \\
$^{13}$C NMR & Shielding-distribution Wasserstein distance & 0.10 ppm \\
IR & Peak-overlap distance & 0.05 \\
Raman & Peak-overlap distance & 0.05 \\
UV--Vis & Excitation-energy Wasserstein distance & 0.08 eV \\
\bottomrule
\end{tabular}
\end{table}

NMR filtering uses scaffold-only computed shieldings, with integration for $^1$H and degeneracy for $^{13}$C. Wasserstein distances compare normalized empirical distributions. For IR and Raman, peak heights are normalized to unit total mass and matched one-to-one within a 20~cm$^{-1}$ position tolerance; the distance is one minus the maximum matched mass, with each match contributing the smaller of the two peak masses. UV--Vis filtering uses the unweighted energy distribution of the first ten excitation states. The alternative NMR representations are evaluated on the same selected families rather than filtered independently. These channel-specific numerical criteria define the curated cohort.

\subsection{Molecular Identity and Spectral Alignment}
The mapping uses molecular derivative data to reconstruct IR and Raman profiles, searches their respective full tables independently, and requires agreement on the spectral row. Mapping acceptance requires IR cosine similarity at least 0.99, Raman similarity at least 0.995, and a best-versus-second-best margin of at least 0.001 in each search. File identity, atom counts, geometry checks, and source-file hashes provide additional checks. These thresholds validate identity alignment; they are distinct from the candidate-separation thresholds in Table~\ref{tab:construction-thresholds}. The observations come from the verified spectral rows, not the reconstructed profiles.

\subsection{Coverage and Intervention Identity}
All 136 source tasks have the required target-property inputs. Across the 11 evaluated representations, the cohort contains 1,496 source-property instances and 2,766 eligible CF substitutions. Bundle CF records require all constituent channels to correspond to the same source and CF candidate.

All-properties CF input combines the constituent properties of the same CF candidate, with candidate order and CF labels checked across channels.

\subsection{Input Boundary and Reproducible Artifacts}
Only the public task projection is supplied: record ID, masked scaffold, allowed attachment fields, candidate SMILES, and the selected property information. Source labels, CF-candidate identities, source-data alignment metadata, and internal validation records remain evaluator-only. Models are not given the candidates' full property table. Evaluation identities include dataset version, cohort, view, and record ID.

A fixed cohort manifest records the selected source IDs, candidate families, property-specific criteria, eligible CF IDs, shuffle assignments, and input SHA256 hashes. Validation covers spectrum mappings, molecular identities, task records, candidate order, CF-candidate labels, and model-visible property payloads. All reported statistics must use the selected source IDs and their eligible CF records.

\section{Physical Meaning and Theoretical Background of Molecular Properties}
\label{app:property_background}
Our 11 property channels cover seven property types and describe molecular charge distribution, frontier electronic structure, local nuclear environments, and vibrational and electronic excitations. All values are quantum-chemical calculations from QM9 and its extensions, QM9NMR and QM9S~\citep{ramakrishnan2014qm9,gupta2021qm9nmr,zou2023qm9s}. We summarize the underlying density functional theory (DFT) quantities and their structural information below. Appendix~\ref{app:properties} lists source-specific calculation levels and input representations.

\subsection{Electronic structures under the DFT formulation}
The starting inputs are nuclear charge numbers $Z_A$, isotope masses $M_A$, nuclear positions $\mathbf R_A$, and the electron number and spin state; $A$ indexes atoms. For the neutral molecules considered here, the electron number is $N_e=\sum_A Z_A$. A density functional and an atomic-orbital basis set specify the computational approximation. At fixed nuclear positions and zero applied field, the nonrelativistic, spin-restricted closed-shell generalized Kohn--Sham equations give
\begin{equation}
\label{eq:prop_gks}
\begin{aligned}
\left[-\frac{\hbar^2}{2m_e}\nabla^2+v_{\mathrm{nuc}}(\mathbf r)
+v_{\mathrm H}[n](\mathbf r)+\hat v_{\mathrm{xc}}\right]\phi_p(\mathbf r)
&=\varepsilon_p\phi_p(\mathbf r),\\
n(\mathbf r)&=\sum_p f_p|\phi_p(\mathbf r)|^2.
\end{aligned}
\end{equation}
Here $p$ indexes normalized spatial orbitals $\phi_p$, $\varepsilon_p$ is an orbital energy, and $f_p$ is its occupation (two for an occupied orbital and zero for an unoccupied orbital), with $\sum_p f_p=N_e$. The variable $\mathbf r$ is an electronic position, $n$ is electron density, $\nabla^2$ is the spatial Laplacian, $m_e$ is the electron mass, and $\hbar=h/(2\pi)$ with Planck's constant $h$. The nuclear-attraction and electron-repulsion (Hartree) potentials are
\begin{equation}
\label{eq:prop_potentials}
v_{\mathrm{nuc}}(\mathbf r)=-\frac{e^2}{4\pi\epsilon_0}\sum_A\frac{Z_A}{|\mathbf r-\mathbf R_A|},
\qquad
v_{\mathrm H}[n](\mathbf r)=\frac{e^2}{4\pi\epsilon_0}\int\frac{n(\mathbf r')}{|\mathbf r-\mathbf r'|}\,d\mathbf r'.
\end{equation}
Here $e$ is the positive elementary charge, $\epsilon_0$ is vacuum permittivity, $\mathbf R_A$ is the position of nucleus $A$, and $\mathbf r'$ is an integration coordinate. The exchange--correlation operator $\hat v_{\mathrm{xc}}$ is obtained from the chosen functional; for hybrid DFT it includes nonlocal exchange. Thus, the orbitals and density are solved together, rather than supplied as independent properties.

The DFT electronic energy $E_{\mathrm{el}}$ sums orbital kinetic energy, electron--nuclear attraction, Hartree repulsion, and the chosen exchange--correlation energy. Adding nuclear repulsion gives the Born--Oppenheimer potential energy,
\begin{equation}
\label{eq:prop_energy}
U(\mathbf R)=E_{\mathrm{el}}(\mathbf R)
+\frac{e^2}{4\pi\epsilon_0}\sum_{A<B}\frac{Z_AZ_B}{|\mathbf R_A-\mathbf R_B|},
\qquad \nabla_{\mathbf R}U(\mathbf R^0)=0.
\end{equation}
Here $\mathbf R$ collects all nuclear coordinates, $A<B$ counts each nuclear pair once, and $\mathbf R^0$ is a stationary geometry found by energy minimization; identifying it as a stable minimum additionally requires positive curvature in the vibrational subspace. $\nabla_{\mathbf R}$ differentiates with respect to those coordinates. Electronic properties and their derivatives below are evaluated using the appropriate source's optimized geometry and calculation level. The functional and basis are required model choices, so these approximate properties are not determined by masses and universal constants alone.

\subsection{Electric Dipole Moment}
The electric dipole operator is
\begin{equation}
\label{eq:prop_dipole_operator}
\hat{\boldsymbol\mu}=e\sum_A Z_A\mathbf R_A-e\sum_{\ell=1}^{N_e}\hat{\mathbf r}_{\ell}.
\end{equation}
Here $\ell$ indexes electrons and $\hat{\mathbf r}_{\ell}$ is their position operator; nuclear charges and positions are defined above. The permanent dipole is the ground-state expectation of this operator and measures the separation of positive and negative charge. Using the electron density from Eq.~\eqref{eq:prop_gks}, it is
\begin{equation}
\label{eq:prop_permanent_dipole}
\boldsymbol\mu=e\sum_A Z_A\mathbf R_A-e\int\mathbf r\,n(\mathbf r)\,d\mathbf r,
\qquad \mu=\|\boldsymbol\mu\|.
\end{equation}
Here, the vector $\boldsymbol\mu$ is the permanent dipole moment, and $\mu$ is its magnitude; $\|\cdot\|$ denotes the Euclidean norm. The dipole reflects bond polarity and the geometry-dependent cancellation of bond contributions; even polar bonds can give zero net dipole in a symmetric molecule. It therefore constrains overall polarity, substituent arrangement, and conformation.

\subsection{HOMO--LUMO Gap}
The frontier-orbital gap is the energy separation between the highest occupied molecular orbital (HOMO) and lowest unoccupied molecular orbital (LUMO):
\begin{equation}
\label{eq:prop_gap}
\Delta\varepsilon=\varepsilon_{\mathrm{LUMO}}-\varepsilon_{\mathrm{HOMO}}.
\end{equation}
Here $\varepsilon_{\mathrm{HOMO}}$ and $\varepsilon_{\mathrm{LUMO}}$ are the respective orbital energies, and $\Delta\varepsilon$ is their difference. These are the eigenvalues defined in Eq.~\eqref{eq:prop_gks}. Their separation characterizes frontier electronic structure and is sensitive to conjugation and electron-donating or electron-withdrawing substituents. Within related molecules, extended conjugation often reduces this gap. It is a method-dependent electronic descriptor, generally distinct from both the fundamental gap (ionization energy minus electron affinity) and a neutral excitation energy~\citep{baerends2013gaps}.

\subsection{Proton and Carbon-13 NMR}
The nuclear magnetic shielding tensor describes how the electronic response to an external magnetic field modifies the field at nucleus $i$:
\begin{equation}
\label{eq:prop_shielding}
\mathbf B_{\mathrm{loc},i}=(\mathbf I-\boldsymbol\sigma_i)\mathbf B_{\mathrm{ext}},
\qquad
\sigma_i^{\mathrm{iso}}=\frac{1}{3}\operatorname{Tr}(\boldsymbol\sigma_i).
\end{equation}
Here $i$ indexes nuclei, $\mathbf B_{\mathrm{ext}}$ is the external magnetic field, and $\mathbf B_{\mathrm{loc},i}$ is the local field at nucleus $i$. The tensor $\mathbf I$ is the identity, $\boldsymbol\sigma_i$ is the dimensionless nuclear shielding tensor, and $\sigma_i^{\mathrm{iso}}$ is its isotropic average. The trace $\operatorname{Tr}$ sums the three diagonal tensor components. DFT magnetic-response calculations yield $\boldsymbol\sigma_i$; multiplying its dimensionless isotropic average by $10^6$ gives the numerical value in ppm.

The tensor itself is obtained as a mixed derivative of the electronic energy with respect to magnetic perturbations~\citep{jinger2021shieldings}:
\begin{equation}
\label{eq:prop_magnetic_response}
\sigma_{i,ab}=\left.\frac{\partial^2 E_{\mathrm{el}}(\mathbf B_{\mathrm{ext}},\mathbf m_i;\mathbf R^0)}
{\partial m_{i,a}\,\partial B_{\mathrm{ext},b}}\right|_{\mathbf B_{\mathrm{ext}}=\mathbf m_i=0}.
\end{equation}
Here $a,b\in\{x,y,z\}$ are Cartesian components and $\mathbf m_i$ is the magnetic moment at nucleus $i$. The fields perturb the electronic Hamiltonian; differentiating the self-consistent electronic solution gives the magnetic response at fixed geometry. In this formula $E_{\mathrm{el}}$ excludes the direct nuclear Zeeman term $-\mathbf m_i\cdot\mathbf B_{\mathrm{ext}}$, which accounts for the unshielded field. The denominator has units of energy, explaining why $\sigma_{i,ab}$ is dimensionless.

\paragraph{$^1$H NMR.}
Proton shielding is sensitive to local bonding and electronic effects, including nearby electronegative atoms, aromatic ring currents, and intramolecular hydrogen bonding. It helps distinguish hydrogen environments and substitution patterns, including changes induced at retained hydrogens by the masked group.

\paragraph{$^{13}$C NMR.}
Carbon shielding helps distinguish saturated, unsaturated, aromatic, and carbonyl environments, thereby constraining the carbon framework and functional groups. For both nuclei, we expose only retained-scaffold values. Optional equivalence-class counts encode how many retained nuclei share a computed class; they do not encode spin--spin splitting.

\subsection{Infrared and Raman Vibrational Properties}
Both channels describe normal modes about an optimized geometry. In the harmonic approximation~\citep{wilson1980vibrations}, differentiating the potential energy in Eq.~\eqref{eq:prop_energy} gives the Cartesian Hessian, and the isotope masses give the mass matrix:
\begin{equation}
\label{eq:prop_hessian}
H_{Aa,Bb}=\left.\frac{\partial^2 U}{\partial R_{Aa}\,\partial R_{Bb}}\right|_{\mathbf R^0},
\qquad M_{Aa,Bb}=M_A\delta_{AB}\delta_{ab}.
\end{equation}
Here $A,B$ index atoms, $a,b$ index Cartesian directions, $R_{Aa}$ is a component of $\mathbf R_A$, and each Kronecker delta is one when its indices match and zero otherwise. The normal modes follow by diagonalization:
\begin{equation}
\label{eq:prop_normal_modes}
\mathbf M^{-1/2}\mathbf H\mathbf M^{-1/2}\mathbf L_j=\omega_j^2\mathbf L_j,
\qquad
\widetilde\nu_j=\frac{\omega_j}{2\pi c_0},
\end{equation}
Here $j$ indexes vibrational modes, $\mathbf L_j$ is the dimensionless eigenvector of the mass-weighted Hessian for mode $j$, $\omega_j$ is its angular frequency, $\widetilde\nu_j$ is its wavenumber, and $c_0$ is the speed of light. Each atomic mass appears three times on the diagonal of $\mathbf M$, once per Cartesian coordinate; $\mathbf M^{-1/2}$ contains their inverse square roots. Frequencies depend on atomic masses, bond stiffness, and coupling between atomic motions.

In a fixed molecular frame, remove overall translation and rotation and choose orthonormal eigenvectors at $\mathbf R^0$, $\mathbf L_j^{\mathsf T}\mathbf L_k=\delta_{jk}$, where $k$ is another mode index and $\mathsf T$ denotes transpose. The mass-weighted normal coordinate is then explicitly
\begin{equation}
\label{eq:prop_normal_coordinates}
\begin{aligned}
Q_j&=\mathbf L_j^{\mathsf T}\mathbf M^{1/2}(\mathbf R-\mathbf R^0)
=\sum_{A,a}L_{Aa,j}\sqrt{M_A}\,(R_{Aa}-R^0_{Aa}),\\
\frac{\partial}{\partial Q_j}&=\sum_{A,a}\frac{L_{Aa,j}}{\sqrt{M_A}}\frac{\partial}{\partial R_{Aa}}.
\end{aligned}
\end{equation}
Here $L_{Aa,j}$ is the component of $\mathbf L_j$ for atom $A$ and direction $a$. Thus $Q_j$ is a projection of mass-weighted atomic displacements, with units of square root of mass times length, rather than a new independent input. The second relation uses these fixed reference eigenvectors and converts Cartesian property derivatives to mode derivatives with the other mode coordinates held fixed. These derivatives may be evaluated by analytic response equations or by finite differences of self-consistent calculations at displaced geometries.

\paragraph{Infrared (IR).}
In the double-harmonic approximation (harmonic nuclear motion and a dipole linear in displacement), the IR intensity of mode $j$ is proportional to the squared dipole derivative along its normal coordinate $Q_j$:
\begin{equation}
\label{eq:prop_ir}
I_j^{\mathrm{IR}}\propto\left\|\left(\frac{\partial\boldsymbol\mu}{\partial Q_j}\right)_0\right\|^2.
\end{equation}
Here $I_j^{\mathrm{IR}}$ is the integrated harmonic IR intensity of mode $j$. The derivative $\partial\boldsymbol\mu/\partial Q_j$ is the change in the permanent dipole per unit normal-coordinate displacement; the subscript $0$ denotes evaluation at the equilibrium geometry. IR activity therefore requires a vibration to change the dipole. Band positions constrain functional groups such as carbonyl, O--H, and N--H groups, while the fingerprint region reflects coupled skeletal vibrations.

\paragraph{Raman.}
Raman activity instead arises from changes in the polarizability tensor $\boldsymbol\alpha$, the response of the dipole to an electric field. Its mode derivative must be nonzero, $(\partial\boldsymbol\alpha/\partial Q_j)_0\ne0$. Again, the subscript $0$ denotes equilibrium geometry, and the derivative describes the change in polarizability per unit normal-coordinate displacement. Its isotropic and anisotropic parts determine the Raman activity. Raman often emphasizes symmetric skeletal and conjugated-bond vibrations. Thus, IR and Raman share normal-mode frequencies but provide complementary intensity patterns.

For the usual static, nonresonant approximation, first define the field-dependent dipole and then differentiate it at zero field:
\begin{equation}
\label{eq:prop_electric_response}
\begin{aligned}
\mu_a(\mathbf R,\mathbf E)&=-\frac{\partial U(\mathbf R,\mathbf E)}{\partial E_a},\\
\alpha_{ab}(\mathbf R)&=\left.\frac{\partial\mu_a(\mathbf R,\mathbf E)}{\partial E_b}\right|_{\mathbf E=0}
=-\left.\frac{\partial^2 U(\mathbf R,\mathbf E)}{\partial E_a\,\partial E_b}\right|_{\mathbf E=0}.
\end{aligned}
\end{equation}
Here $\mathbf E$ is a uniform electric field, $E_a$ and $\mu_a$ are Cartesian components, and $U(\mathbf R,\mathbf E)$ includes the field interaction of both electrons and nuclei at fixed $\mathbf R$. The permanent dipole in Eq.~\eqref{eq:prop_permanent_dipole} is $\boldsymbol\mu(\mathbf R,\mathbf 0)$; the polarizability derivative is taken before setting the field to zero. These are clamped-nuclei electronic derivatives: the electronic state responds to the field while the nuclei remain fixed. Combining them with Eq.~\eqref{eq:prop_normal_coordinates} gives the IR dipole derivative and the Raman polarizability derivative; both ultimately depend on the same electronic calculation, nuclear geometry, and masses.

\subsection{UV--Visible Electronic Excitations}
Time-dependent DFT (TDDFT) describes the electronic response to a weak time-dependent perturbation. Its linear-response equations are built from occupied and unoccupied ground-state orbitals, their energy differences, and the Hartree and exchange--correlation response kernel (the change of the corresponding potential/operator as the electronic state changes). Solving this response problem yields excitation energies and amplitudes; contraction of the amplitudes with orbital dipole integrals yields transition moments. In electronic-state notation, these quantities are
\begin{equation}
\label{eq:prop_electronic_transitions}
\begin{aligned}
\Delta E_s&=E_s-E_0,
\qquad \boldsymbol\mu_{0s}=\langle\Psi_0|\hat{\boldsymbol\mu}|\Psi_s\rangle,\\
f_{0s}&=\frac{2m_e\Delta E_s}{3\hbar^2 e^2}\|\boldsymbol\mu_{0s}\|^2.
\end{aligned}
\end{equation}
Here $0$ denotes the electronic ground state and $s$ indexes excited states. The quantities $E_0$ and $E_s$ are their energies at the same nuclear geometry, and $\Delta E_s$ is the excitation energy. The states $\Psi_0$ and $\Psi_s$ are normalized electronic wavefunctions, $\hat{\boldsymbol\mu}$ is defined in Eq.~\eqref{eq:prop_dipole_operator}, and the bracket denotes its transition matrix element $\boldsymbol\mu_{0s}$, distinct from the permanent dipole. In Eq.~\eqref{eq:prop_electronic_transitions}, $f_{0s}$ is the dimensionless orientation-averaged oscillator strength of the $0\to s$ transition and $\|\boldsymbol\mu_{0s}\|$ is the transition-dipole magnitude. The transition energies and strengths reflect chromophores, conjugation, and the character of excitations. Their joint pattern provides richer electronic information than a single frontier-orbital gap.

\section{Property Representations}
\label{app:properties}
The property information supplied to a model depends on both the computed source and its conversion into the prompt. The following descriptions specify the reported representations and their preprocessing (Appendix~\ref{app:release}).
\paragraph{Electronic gap.}
We use the QM9 HOMO--LUMO gap in Hartree. QM9 reports properties calculated at the B3LYP/6-31G(2df,p) level~\citep{ramakrishnan2014qm9}. This scalar is used as property information for choosing a completion, not as a regression target for the evaluated language model.

\paragraph{Dipole.}
QM9S provides dipole information in Debye. The norm representation contains its scalar magnitude; the vector representation contains its three Cartesian components. A vector depends on the coordinate frame in which it is represented. We treat norm and vector as different property representations.

\paragraph{NMR.}
The NMR channel uses gas-phase isotropic shielding from QM9NMR, calculated at the mPW1PW91/6-311+G(2d,p) level~\citep{gupta2021qm9nmr}. Values are shielding constants in ppm, not experimentally referenced chemical shifts. Only nuclei belonging to the retained scaffold are exposed. Symmetry classes are determined in the completed molecule, so a hidden-group change can alter both a retained nucleus's shielding and its equivalence class. The peaks-only view reports class-averaged shieldings. The views labeled integration ($^1$H) or degeneracy ($^{13}$C) also report retained-nucleus counts for these equivalence classes. These are computed counts, not measured peak areas or spin--spin splitting multiplicities. The raw view contains one value per retained nucleus without internal atom identifiers. These views probe representation sensitivity without changing the molecular candidate set.

\paragraph{IR and Raman.}
The spectra are derived from QM9S calculations at the B3LYP/def-TZVP level~\citep{zou2023qm9s}. The supplied property information is a compact peak list from the broadened spectrum rather than a raster plot or its full 3,501-point grid. Wavenumbers are in $\mathrm{cm}^{-1}$ and intensities are relative. Peak extraction uses relative-prominence thresholds of 0.02 for IR and 0.03 for Raman.

\paragraph{UV--Vis.}
The UV--Vis view uses the first ten QM9S TDDFT transitions, exposing excitation energy in eV, wavelength in nm, and transition-dipole norm in Debye.

\paragraph{Combined properties.}
The all-properties condition concatenates the seven canonical channels: dipole norm, gap, $^1$H NMR peaks with retained-nucleus counts, $^{13}$C NMR peaks with degeneracy, IR, Raman, and UV--Vis.

\subsection{Within-Family Property Separability}
\label{app:channel-separability}
Table~\ref{tab:channel-separability} summarizes the 25 September 2026 analysis of the 136 final-cohort source tasks drawn from release \texttt{0.1.0-candidate.15-spectra-r1}, using the released source-property and CF-property inputs. A task meets a channel's separation criterion when the source observations fall outside the similarity cutoff of every distractor with an available released input. The percentage is calculated over tasks with at least one available distractor, not over candidate pairs.

\noindent\textbf{Electronic properties.} Dipole magnitude and gap use absolute differences, with separation cutoffs of $0.5$ D and $0.1$ eV, respectively. Gap values are converted from Hartree using $27.211386245988$ eV per Hartree. UV--Vis compares the first ten excitation states in their supplied order. For a source--distractor pair, let $d_E$ and $d_\mu$ be the largest absolute differences across these states in excitation energy and transition-dipole norm. The normalized distance is $\max\{d_E/(0.2\,\mathrm{eV}),d_\mu/(0.25\,\mathrm{D})\}$, with a pair inside the similarity cutoff when this distance is at most one. The smallest nearest-distractor distance across the evaluated tasks is 1.94, and the median is 6.49.

\noindent\textbf{Vibrational properties.} IR and Raman use the released peak lists. Nonnegative peak intensities are normalized to sum to one within each spectrum. A one-to-one peak assignment maximizes matched mass, with each pair contributing the smaller normalized intensity only when the wavenumber difference is at most $10\,\mathrm{cm}^{-1}$. Let $M$ denote the total matched mass. A distractor is inside the similarity cutoff if $M\geq0.9$; the distance reported in the table is $1-M$. Larger distances indicate less overlap of the supplied peak lists.

\noindent\textbf{NMR representations.} Shielding observations are treated as unordered multisets. Peaks-only inputs contribute one value per peak; integration- and degeneracy-augmented inputs repeat each value according to its supplied count; atomwise inputs use each listed value once. For equal-sized multisets, a one-to-one assignment minimizes the sum of absolute shielding differences. A pair is inside the similarity cutoff when all assigned differences are at most $0.1$ ppm for $^1$H or $2$ ppm for $^{13}$C. The pairwise distance is the largest difference under that assignment. Unequal multiset sizes are treated as a mismatch. Distance summaries use tasks with finite nearest-distractor distances.

\begin{table}[htbp]
\centering
\small
\setlength{\tabcolsep}{4pt}
\caption{\textbf{Channel-wise property separability.} Percentages refer to source tasks separated from every available distractor under the channel-specific reporting cutoffs described above. Distance summaries use the nearest available distractor for each task; Q1--Q3 is the interquartile range. Dipole distances are in D, gap in eV, and NMR in ppm; UV--Vis distances are cutoff-normalized, while IR and Raman use $1-M$.}
\label{tab:channel-separability}
\begin{tabular}{lrrr}
\toprule
\textbf{Channel} & \textbf{\%} & \textbf{Median} & \textbf{Q1--Q3}\\
\midrule
Dipole magnitude & 95.4 & 1.184 & 0.806--1.566\\
HOMO--LUMO gap & 88.2 & 0.912 & 0.403--1.510\\
UV--Vis & 100.0 & 6.486 & 5.309--8.980\\
IR & 100.0 & 0.713 & 0.618--0.796\\
Raman & 99.2 & 0.435 & 0.312--0.567\\
\midrule
$^1$H peaks & 94.1 & 0.808 & 0.291--1.342\\
$^1$H peaks + integration & 94.1 & 0.857 & 0.324--1.377\\
$^1$H atomwise & 99.3 & 0.922 & 0.416--1.427\\
$^{13}$C peaks & 96.3 & 11.438 & 6.782--25.128\\
$^{13}$C peaks + degeneracy & 96.3 & 11.897 & 6.829--24.983\\
$^{13}$C atomwise & 98.5 & 11.718 & 6.829--24.983\\
\bottomrule
\end{tabular}
\end{table}

The final-cohort analysis is implemented in \path{audits/final136_appendix_sync/recompute.py}, reusing the reporting rules in \path{audits/current/scripts/analyze_channel_separability.py}. Its archived summary records the channel counts, distance quantiles, dataset version, and SHA256 fingerprints of the source and CF inputs used for the calculation.

\section{Inference Configuration and Coverage}
\label{app:inference}
\subsection{Model Identifiers and Protocol}
The seven evaluated models are recorded as \texttt{gpt-5.6-luna}, \texttt{gpt-5.6-sol}, \texttt{gpt-6-astra}, \texttt{claude-opus-5}, \texttt{Qwen/Qwen3.8-27B}, \texttt{google/gemma-4-31B-it}, and the local deployment alias \texttt{chemdfm-v2:14b-f16}.

No tools or literature retrieval are enabled.

\subsection{Coverage and Output Selection}
Models return a zero-based candidate index and a concise justification. Source-property accuracy uses all 136 tasks. Paired source--shuffle comparisons use input-validated matched records. Appendix~\ref{app:cf} defines CF eligibility and source weighting.

Scoring selects the first terminal occurrence for the applicable versioned input. API failures can be retried; a terminal invalid answer is retained and scored as incorrect, not replaced by a later convenient response. Ordinary accuracy and unconditional CF accuracy include invalid outputs in their full denominators. CFR instead requires a correct source prediction and valid responses to eligible CF substitutions.

\subsection{Bootstrap Intervals}
We retain task-level predictions for paired model comparisons. Where reported, 95\% percentile intervals use paired source-task bootstrap resamples.

The intervals in Table~\ref{tab:combined} use 10,000 paired bootstrap resamples of the 136 source-task correctness differences, with seed 20260916 and percentile endpoints. Invalid answers are incorrect. Source tasks are the resampling units.

\section{Additional Source-Recovery and Representation Results}
\label{app:repr}
Table~\ref{tab:representations} supplies the broader model and representation context. Every reported value uses the same 136 source tasks. NMR values are computed scaffold shieldings; changing their representation changes retained information as well as presentation. The canonical-channel and bundle rows use the same corrected snapshot as the main-text tables; the additional representation rows are recomputed on the same final cohort after exact input replay.

\begin{table}[htbp]
\centering\scriptsize
\setlength{\tabcolsep}{3pt}
\caption{Source-property accuracy (\%) across seven models. L: Luna; S: Sol; A: Astra; O: Opus; Q: Qwen; G: Gemma; C: ChemDFM. All denominators are 136, including invalid responses.}
\label{tab:representations}
\begin{tabular*}{\linewidth}{@{\extracolsep{\fill}}lrrrrrrr@{}}
\toprule
Property & L & S & A & O & Q & G & C\\
\midrule
No properties ($P_0$) & 39.0 & 42.6 & 41.2 & 33.1 & 33.8 & 38.2 & 39.0\\
Dipole norm & 65.4 & 63.2 & 69.1 & 60.3 & 44.9 & 44.9 & 39.0\\
Dipole vector & 62.5 & 69.1 & 71.3 & 61.8 & 37.5 & 40.4 & 38.2\\
Gap & 41.9 & 71.3 & 69.1 & 53.7 & 36.8 & 41.2 & 36.0\\
$^1$H peaks & 37.5 & 54.4 & 69.9 & 55.1 & 32.4 & 39.7 & 39.0\\
$^1$H +Int. & 44.9 & 61.0 & 69.9 & 55.1 & 35.3 & 38.2 & 41.9\\
$^1$H atomwise & 44.1 & 58.1 & 65.4 & 63.2 & 36.0 & 38.2 & 40.4\\
$^{13}$C peaks & 48.5 & 54.4 & 66.2 & 69.1 & 34.6 & 33.8 & 34.6\\
$^{13}$C +Deg. & 49.3 & 55.1 & 67.6 & 66.2 & 36.0 & 36.0 & 36.0\\
$^{13}$C atomwise & 54.4 & 52.2 & 66.9 & 70.6 & 39.0 & 36.8 & 36.0\\
IR & 81.6 & 89.0 & 93.4 & 88.2 & 55.1 & 69.9 & 40.4\\
Raman & 69.1 & 77.9 & 80.9 & 77.2 & 51.5 & 61.8 & 40.4\\
UV--Vis & 53.7 & 72.8 & 80.1 & 67.6 & 36.0 & 44.9 & 36.0\\
\midrule
Electronic & 61.8 & 73.5 & 88.2 & 68.4 & 36.0 & 44.9 & 40.4\\
NMR & 44.1 & 62.5 & 76.5 & 55.1 & 37.5 & 39.0 & 41.2\\
Vibrational & 83.8 & 89.7 & 96.3 & 90.4 & 64.0 & 74.3 & 40.4\\
All properties & 78.7 & 93.4 & 98.5 & 89.0 & 48.5 & 54.4 & 39.7\\
\bottomrule
\end{tabular*}
\end{table}

\section{Property Evaluation: Coverage, Counterfactual Flips, and Metrics}
\label{app:cf}

\subsection{Conditional Cohorts and Source Weighting}
Table~\ref{tab:cf-transition} reports Counterfactual Flip Rates (CFR) for all seven models. A source enters only when the model selects it correctly under source properties and has at least one valid response to an eligible CF substitution. After the same input-replay exclusions as in the main evaluation, valid CF responses are averaged within source and then across sources (Eq.~\ref{eq:transition}). Table~\ref{tab:cf} additionally separates successful CF flips, source persistence, and other choices for selected spectral and combined inputs.

\begin{table}[htbp]
\centering\small
\setlength{\tabcolsep}{4pt}
\caption{Conditional source-macro transitions (\%). The CFR column uses Eq.~\ref{eq:transition}; source-persistence and other-choice rates use the same cohort and source weighting with the corresponding outcome indicators. Minor sum differences reflect rounding.}
\label{tab:cf}
\begin{tabular}{llrrr}
\toprule
Model & Property & CFR & Persist & Other\\
\midrule
Luna & IR & 80.3 & 11.1 & 8.6\\
 & Raman & 56.2 & 27.9 & 15.9\\
 & Vibrational & 85.0 & 8.5 & 6.5\\
 & All & 75.8 & 16.2 & 8.0\\
\midrule
Sol & IR & 90.1 & 6.2 & 3.6\\
 & Raman & 63.8 & 25.0 & 11.2\\
 & Vibrational & 90.2 & 5.6 & 4.2\\
 & All & 88.1 & 7.9 & 4.0\\
\midrule
Astra & IR & 91.2 & 5.1 & 3.7\\
 & Raman & 65.1 & 23.8 & 11.1\\
 & Vibrational & 92.1 & 5.7 & 2.1\\
 & All & 97.9 & 1.7 & 0.4\\
\midrule
Opus & IR & 89.8 & 5.5 & 4.7\\
 & Raman & 58.7 & 27.7 & 13.7\\
 & Vibrational & 89.3 & 6.3 & 4.4\\
 & All & 83.6 & 10.3 & 6.1\\
\bottomrule
\end{tabular}
\end{table}

\subsection{Unconditional CF Accuracy}
Table~\ref{tab:counterfactual-recovery} reports unconditional CF accuracy on the final cohort. Unlike Table~\ref{tab:cf-transition}, this pair-micro metric includes initially incorrect sources and counts terminal-invalid CF responses as failures.

\begin{table}[htbp]
\caption{\textbf{Unconditional CF Accuracy.} CF-candidate selection accuracy (\%) under same-scaffold property substitution, without conditioning on source correctness. Eligible, input-validated CF pairs receive equal weight; terminal-invalid responses count as failures. Model identifiers and property definitions follow Table~\ref{tab:evidence-pairing-gain}. Green shading and boldface mark the highest value in each column.}
\label{tab:counterfactual-recovery}
\begin{center}
\begingroup
\setlength{\tabcolsep}{3.2pt}
\renewcommand{\arraystretch}{1.15}
\small
\resizebox{\linewidth}{!}{%
\begin{tabular}{l*{11}{r}}
\toprule
 & \multicolumn{3}{c}{\textbf{Electronic}} & \multicolumn{2}{c}{\textbf{Vibrational}} & \multicolumn{6}{c}{\textbf{Magnetic Resonance}} \\
\cmidrule(lr){2-4}\cmidrule(lr){5-6}\cmidrule(lr){7-12}
 & & & & & & \multicolumn{3}{c}{$^1$H NMR} & \multicolumn{3}{c}{$^{13}$C NMR} \\
\cmidrule(lr){7-9}\cmidrule(lr){10-12}
\textbf{Model} & Dipole & Gap & UV--Vis & IR & Raman & Peaks & +Int. & Atomwise & Peaks & +Deg. & Atomwise \\
\midrule
\multicolumn{12}{l}{\itshape Closed-weight LLMs} \\
\midrule
GPT-5.6-Luna & $54.2$ & $45.7$ & $60.0$ & $80.4$ & $61.7$ & $33.7$ & $37.6$ & $32.2$ & $38.4$ & $31.0$ & $32.2$ \\
\rowcolor{black!4}
GPT-5.6-Sol & $60.0$ & $65.9$ & $70.0$ & $90.0$ & $66.7$ & $50.0$ & $46.9$ & $48.4$ & $56.2$ & $53.9$ & $54.3$ \\
GPT-6-Astra & \cellcolor{green!12}\textbf{66.7} & \cellcolor{green!12}\textbf{70.2} & \cellcolor{green!12}\textbf{77.9} & \cellcolor{green!12}\textbf{90.4} & \cellcolor{green!12}\textbf{70.8} & \cellcolor{green!12}\textbf{64.7} & \cellcolor{green!12}\textbf{66.3} & \cellcolor{green!12}\textbf{57.8} & \cellcolor{green!12}\textbf{65.9} & \cellcolor{green!12}\textbf{67.4} & \cellcolor{green!12}\textbf{65.1} \\
\rowcolor{black!4}
Claude Opus 5 & $52.5$ & $52.7$ & $62.1$ & $86.7$ & $66.2$ & $50.8$ & $50.0$ & $48.4$ & $63.6$ & $58.5$ & $64.0$ \\
\midrule
\multicolumn{12}{l}{\itshape Open-weight LLMs} \\
\midrule
Qwen3.8-27B & $37.1$ & $32.2$ & $31.2$ & $53.3$ & $43.8$ & $33.1$ & $31.9$ & $33.5$ & $35.0$ & $33.1$ & $34.5$ \\
\rowcolor{black!4}
Gemma-4-31B-it & $40.4$ & $30.2$ & $41.7$ & $63.8$ & $45.8$ & $31.8$ & $31.4$ & $32.2$ & $34.9$ & $32.2$ & $34.5$ \\
ChemDFM-v2.0-14B & $30.4$ & $34.1$ & $32.1$ & $33.3$ & $30.8$ & $31.8$ & $29.1$ & $35.3$ & $34.1$ & $31.8$ & $32.9$ \\
\bottomrule
\end{tabular}%
}
\endgroup
\end{center}
\end{table}

\subsection{Comparison of Property Rankings}
\label{app:pairing-transition-bootstrap}
We examine the atomwise $^1$H versus $^{13}$C peaks contrast highlighted in Section~\ref{sec:results}. Luna, Sol, and Astra each have a larger Reliance Margin (RM) for atomwise $^1$H, but a lower CFR (Tables~\ref{tab:evidence-pairing-gain} and~\ref{tab:cf-transition}).

For Sol, RM values are 25.19 and 12.59 pp, respectively, on matched input-validated sources. CFR values are 53.59\% and 67.34\%, respectively. We draw 100,000 paired bootstrap samples of 136 sources with replacement from the full final-cohort source universe (seed 20260922; NumPy 1.26.4, PCG64; lexicographically sorted source IDs). The two views and both metrics share the same sampled source IDs and multiplicities. Each source retains all its CF records; metric-specific eligibility and denominators are recomputed in every replicate, preserving within-source averaging and source-macro weighting. No replicate has an empty denominator.

We define each difference as atomwise $^1$H minus $^{13}$C peaks. The RM difference is $+12.59$ pp (95\% percentile interval: $[-1.50,27.07]$ pp), and the CFR difference is $-13.76$ pp ($[-26.88,-0.56]$ pp). Both directions are preserved in 93,319 of 100,000 replicates (93.3\%). As a sensitivity check, restricting both conditional rates to the sources initially correct in both views and with valid CF responses yields 60.33\% versus 73.33\%, a difference of $-13.00$ pp ($[-28.00,2.00]$ pp). This common-cohort comparison uses a separate paired source bootstrap with the same seed.

\subsection{Metric Definitions}
\label{app:metric-definitions}
We use the task and prediction notation defined in Section~\ref{sec:protocol}. Let $\mathbf 1\{\cdot\}$ denote the indicator function. For source-property and shuffled-property inputs, define
\begin{equation}
 c_{i,e}^{(p)}=\mathbf 1\{\widehat s_{i,e}^{(p)}=s_i\},\qquad e\in\{\mathrm{src},\mathrm{shuf}\},
\end{equation}
and let $c_{i,0}=\mathbf 1\{\widehat s_{i,0}=s_i\}$. Terminal-invalid responses count as incorrect in these accuracy calculations.

\noindent\textbf{Reliance Margin (RM).} For each comparison, we retain tasks whose source and shuffled inputs both pass input validation and index these matched source tasks by $i=1,\ldots,N$. Thus, $N$ denotes the number of evaluated source tasks in that comparison, after any input-validation exclusions. RM is
\begin{equation}
 \mathrm{RM}^{(p)}=\frac{100}{N}\sum_{i=1}^{N}\left(c_{i,\mathrm{src}}^{(p)}-c_{i,\mathrm{shuf}}^{(p)}\right).
 \label{eq:pairing-gain}
\end{equation}
Both conditions are scored against the source candidate. Table~\ref{tab:evidence-pairing-gain} reports RM in pp. Unlike the comparison with $P_0$, RM retains property inputs in both conditions and tests their correct pairing with the structural task.

\noindent\textbf{Counterfactual Flip Rate (CFR).} Let $J_i^{(p)}$ contain the eligible CF candidates whose CF inputs pass validation and whose model responses yield valid candidate indices. Define the initially correct cohort
\begin{equation}
 S_p=\{i:\widehat s_{i,\mathrm{src}}^{(p)}=s_i,\ |J_i^{(p)}|>0\}.
\end{equation}
CFR, expressed as a percentage, is
\begin{equation}
 \mathrm{CFR}^{(p)}=\frac{100}{|S_p|}\sum_{i\in S_p}\frac{1}{|J_i^{(p)}|}\sum_{j\in J_i^{(p)}}\mathbf 1\{\widehat s_{ij,\mathrm{CF}}^{(p)}=j\}.
 \label{eq:transition}
\end{equation}
We first calculate each task's fraction of valid CF responses selecting the corresponding CF candidate, then average across tasks with equal weight. The source-property response determines cohort inclusion but is not counted as a CF test. Selecting the source or another candidate counts as an unsuccessful flip; an invalid CF response is excluded from this conditional metric. Table~\ref{tab:cf-transition} reports CFR in \%. Empty conditional cohorts yield N/A. The initially correct cohort is defined separately for each model and configuration. In contrast, unconditional CF accuracy averages correctness over all input-validated source--CF candidate pairs and includes terminal-invalid responses as failures.

\noindent\textbf{Auxiliary accuracy and baseline gain.} Source-property accuracy, also termed source recovery, uses the input-validated source cohort of size $N_{\mathrm{src}}$:
\begin{equation}
 A_{\mathrm{src}}^{(p)}=\frac{1}{N_{\mathrm{src}}}\sum_{i=1}^{N_{\mathrm{src}}}c_{i,\mathrm{src}}^{(p)}.
 \label{eq:source-recovery}
\end{equation}
The source-recovery tables use all 136 final-cohort tasks, whereas RM uses the source--shuffle input-valid intersection. On the common source/no-properties cohort, which contains the same 136 tasks here, improvement over the baseline $P_0$ is
\begin{equation}
 \Delta_{0}^{(p)}=\frac{100}{N_{\mathrm{src}}}\sum_{i=1}^{N_{\mathrm{src}}}\left(c_{i,\mathrm{src}}^{(p)}-c_{i,0}\right).
 \label{eq:source-recovery-gain}
\end{equation}
Reported in pp, this source recovery gain measures whether source properties improve recovery beyond the scaffold and candidate structures alone. Uniform random candidate selection has expected accuracy $N_{\mathrm{src}}^{-1}\sum_i n_i^{-1}=37.75\%$ on the final cohort.

\subsection{Model-Ranking Stability under Family Subsampling}
\label{app:family-subsampling}
We assess how aggregate model rankings change when fewer task families are evaluated. For each subset size $n\in\{25,50,75,100,125\}$, we sample 1,000 subsets without replacement from the final cohort (seed 20260925). Each draw uses the same families across all seven models and retains their source, shuffled, and CF records together. Predictions and property assignments remain fixed; only the evaluated family subset changes.

We aggregate over eleven input configurations: the seven single-property channels (using peaks-only NMR), the electronic, magnetic-resonance, and vibrational bundles, and all properties. For each configuration, RM and CFR follow Appendix~\ref{app:metric-definitions}; eligible families receive equal weight, with CF responses first averaged within family. Configuration scores are then averaged equally. RM excludes input-mismatched shuffle records, and CFR includes only initially recovered sources with valid, input-matched CF responses. If a sampled configuration has no eligible family for any model, it is omitted for all models in that draw.

\begin{table}[htbp]
\caption{Model-ranking stability under family subsampling. For each score, $\rho$ is the mean Spearman correlation with its full-cohort model ranking; Exact is the percentage of draws preserving the complete ranking; MAD is the mean absolute score deviation from the full-cohort value, averaged across models and draws, in percentage points.}
\label{tab:family-subsampling}
\begin{center}
\small
\setlength{\tabcolsep}{4pt}
\begin{tabular}{r*{6}{r}}
\toprule
 & \multicolumn{3}{c}{RM} & \multicolumn{3}{c}{CFR} \\
\cmidrule(lr){2-4}\cmidrule(lr){5-7}
$n$ & $\rho$ & Exact & MAD & $\rho$ & Exact & MAD \\
\midrule
25  & 0.9845 & 59.9 & 3.16 & 0.9859 & 63.3 & 2.94 \\
50  & 0.9938 & 82.1 & 2.01 & 0.9936 & 82.2 & 1.80 \\
75  & 0.9981 & 94.5 & 1.40 & 0.9969 & 91.2 & 1.26 \\
100 & 0.9998 & 99.3 & 0.91 & 0.9992 & 97.7 & 0.80 \\
125 & 1.0000 & 100.0 & 0.45 & 1.0000 & 100.0 & 0.41 \\
\bottomrule
\end{tabular}
\end{center}
\end{table}

The full-cohort ordering is Astra, Sol, Opus, Luna, Gemma, Qwen, and ChemDFM for both aggregate metrics. Table~\ref{tab:family-subsampling} shows that mean rank correlations exceed 0.98 even at the smallest subset size. At $n=100$, the complete RM and CFR rankings are retained in 99.3\% and 97.7\% of draws, respectively, with mean absolute deviations below one percentage point. At $n=125$, both rankings are retained in every draw. These results show that the aggregate model ordering is stable to family subsampling within the evaluated cohort.

\subsection{Channel-Wise PGR Stability under Family Subsampling}
\label{app:pgr-subsampling}
We evaluate PGR stability separately for the eleven single-property channels in Table~\ref{tab:pgr-leaderboard}, using the same complete-coverage, common-input-valid family sets and the score in Eq.~\ref{eq:pgr}. We draw 1,000 subsets of 100 families without replacement from the final 136-family cohort, using lexicographically sorted family IDs and NumPy 1.26.4's PCG64 generator with seed 20260925. Each subset is shared across channels and models; for channel $p$, scoring uses its intersection with $\mathcal E_p$. All source and CF outcomes remain paired within family. No predictions are regenerated, and channels are not averaged.

For each channel, Spearman correlation compares the seven-model ranking in each subset with that channel's full-set ranking. Ties receive average ranks. Exact agreement requires the same ranks, including ties. We calculate family scores in integer twelfths before ranking to avoid floating-point tie artifacts. Mean absolute deviation (MAD) averages absolute PGR differences over models and draws. These channel-wise PGR results are separate from the aggregate RM/CFR analysis above.

\begin{table}[htbp]
\centering
\caption{Channel-wise PGR stability over 1,000 subsamples of 100 families. $\rho$ is mean Spearman rank correlation; Exact is complete-ranking agreement (\%); MAD is mean absolute score deviation (pp).}
\label{tab:pgr-subsampling}
\small
\begin{tabular}{lrrr}
\toprule
Property & $\rho$ & Exact & MAD \\
\midrule
Dipole & 0.9865 & 62.4 & 1.34 \\
Gap & 0.9793 & 52.3 & 1.04 \\
UV--Vis & 0.9989 & 96.2 & 1.14 \\
IR & 0.9991 & 97.0 & 0.91 \\
Raman & 0.9887 & 66.1 & 1.08 \\
$^1$H peaks & 0.9398 & 20.0 & 1.10 \\
$^1$H +Int. & 0.9562 & 29.9 & 1.08 \\
$^1$H atomwise & 0.9477 & 23.5 & 1.12 \\
$^{13}$C peaks & 0.9711 & 38.2 & 1.08 \\
$^{13}$C +Deg. & 0.9627 & 34.7 & 1.10 \\
$^{13}$C atomwise & 0.9843 & 61.1 & 1.05 \\
\bottomrule
\end{tabular}
\end{table}

Mean rank correlations range from 0.940 to 0.999, with MAD from 0.91 to 1.34 pp. Complete-ranking agreement varies by channel, from 20.0\% to 97.0\%.

\section{Combined Properties}
\label{app:combined}
The three bundles group related channels: electronic uses dipole norm, gap, and UV--Vis; NMR uses integration-augmented $^1$H and degeneracy-augmented $^{13}$C observations from the visible scaffold; vibrational uses IR and Raman peak lists. All-properties input orders the channels as dipole, gap, $^1$H NMR, $^{13}$C NMR, IR, Raman, and UV--Vis. Table~\ref{tab:representations} reports source accuracy across input combinations; Table~\ref{tab:combined} reports the four paired correctness cells for all-properties versus vibrational input, including invalid outputs as incorrect.

A different reference gives a different contrast. Relative to IR alone, all-properties input gains eight tasks and loses one for Astra (126 correct in both and one wrong in both). Luna gains 12 and loses 16, Sol gains 12 and loses six, and Opus gains 11 and loses ten. This distinguishes marginal score changes from turnover in the error set.

\begin{table}[!htbp]
\caption{Paired source correctness when moving from the vibrational bundle to all-properties input. Each row totals 136. Gain/loss refer to all-properties input relative to vibrational input; loss denotes correct-to-incorrect corruption. These are within-model input comparisons, not cross-model transitions. Intervals are descriptive paired-bootstrap 95\% intervals in pp.}
\label{tab:combined}
\begin{center}
\small
\begin{tabular}{lrrrrr}
\toprule
Model & Both correct & Gain & Loss & Both wrong & $\Delta$ [95\% interval]\\
\midrule
Luna & 98 & 9 & 16 & 13 & $-5.1$ [$-12.5$, $2.2$]\\
Sol & 119 & 8 & 3 & 6 & $3.7$ [$-0.7$, $8.1$]\\
Astra & 130 & 4 & 1 & 1 & $2.2$ [$-0.7$, $5.9$]\\
Opus & 115 & 6 & 8 & 7 & $-1.5$ [$-6.6$, $3.7$]\\
\bottomrule
\end{tabular}
\end{center}
\end{table}

\subsection{Matched Input Conditions}
\label{app:combination-levels}
Table~\ref{tab:representations} in Appendix~\ref{app:repr} reports single-property, bundle, and all-properties source accuracy on the same 136 tasks. Constituent property values, masked scaffolds, and candidate order are matched across conditions. The NMR bundle uses integration-augmented $^1$H and degeneracy-augmented $^{13}$C inputs; All-properties input combines electronic, vibrational, and NMR properties using these two NMR representations.

\FloatBarrier

\section{Case Records and Property Interpretation}
\label{app:cases}
Task references link to the corresponding structures and candidate lists in Appendix~\ref{app:task-families}.
\subsection{Verified Residual-Error Records}
The two Astra all-properties source errors are exhaustively documented here. For task \hyperref[app:family-df6b46bd1891d2d1]{DF6B46BD1891D2D1}, the masked scaffold is \texttt{[*:1]C(C)(C)OC=O}; the supplied candidates, in order, are \texttt{CC(C)(OC=O)C(N)=O}, \texttt{CC(C)(C)OC=O}, \texttt{CC(C)(C=O)OC=O}, and \texttt{CC(=O)C(C)(C)OC=O}. The target index is 1; the joint answer is 3 while all three vibrational configurations answer 1. For task \hyperref[app:family-a7c3c4816799e0c2]{A7C3C4816799E0C2}, the candidates are \texttt{N\#CC(CO)CO}, \texttt{CC(=O)C(CO)CO}, \texttt{COC(CO)CO}, and \texttt{OCC(O)CO}. The target is 3; the four spectral/joint configurations all answer 2. These indices and transitions are saved-output observations.

\subsection{Reference-Geometry Checks for Electronic and NMR Explanations}
\label{app:geometry-cases}
These supplementary cases complement Section~\ref{sec:case-study} and were selected after inspecting errors. We verified each featured Astra user request against the current public input and each submitted answer against its saved raw response. Candidate geometry was read from the native property source, converted to a molecular graph, and checked against the corresponding candidate SMILES. No new conformer search or quantum calculation was performed. Atom indices below are zero-based in the source geometry file.

\paragraph{Dipole case.}
For task \hyperref[app:family-ae2202d59dbd5246]{AE2202D59DBD5246}, the masked scaffold is \path{[*:1]N(C=O)C=O}. Target index 3 is \path{CN(C=O)C=O} (QM9 \texttt{gdb\_341}); Astra selects index 0, \path{O=CN(C=O)C=O} (\texttt{gdb\_2501}). Reading the native QM9S files confirms dipole magnitudes 0.73125 and 3.46903~D, respectively. The latter conformer's directed C$\rightarrow$O axes have pairwise angles $118.5547^\circ$, $14.5898^\circ$, and $133.1442^\circ$, inconsistent with the balanced threefold arrangement invoked in the explanation. These axes diagnose geometry; they are not assumed to be additive electronic dipole contributions. The target--selected-candidate mismatch is directly checked. Luna and Opus choose the same alternative with similar symmetry arguments.

\paragraph{NMR case.}
For task \hyperref[app:family-252307dc6fe392bf]{252307DC6FE392BF}, the scaffold is \path{[*:1]C1(C(N)=O)CC1}. Target index 2 is \path{NC(=O)C1(C(N)=O)CC1} (\texttt{gdb\_54299}); Astra selects index 0, \path{CC(=O)C1(C(N)=O)CC1} (\texttt{gdb\_54290}). The target's retained NH shieldings are 22.2110 and 26.9237~ppm, whereas the selected candidate's corresponding values are 27.3063 and 27.3415~ppm. In the target's QM9NMR reference geometry, retained N7--H15 lies 1.8553~\AA\ from O2 of the other carboxamide, at an N--H$\cdots$O angle of $134.93^\circ$. In the selected candidate, the closest relevant amide H16 lies 3.6631~\AA\ from acetyl O2, at $104.83^\circ$. Astra interprets the unequal retained NH shieldings as evidence for hydrogen bonding with the acetyl carbonyl. Thus the proposed short NH$\cdots$acetyl-O contact is unsupported by the selected reference conformer. Luna, Sol, and Opus select the target in this case, so the limitation is not shared by every model.

\paragraph{Representation and scope.}
The two target NH values above are averaged to 24.56735~ppm with integration two in the symmetry-grouped view. This is graph-class averaging, not a measured conformer-ensemble average. The atomwise view preserves their difference but does not expose atom assignments. Both views omit nuclei in the missing group and contain neither coupling constants nor NOE observations. Dipole and NMR geometry checks use their respective native sources.

\paragraph{Interpretation.}
These analyses therefore highlight a gap between recognizing local chemical motifs and grounding their consequences in candidate-specific geometry. Graph symmetry need not imply dipole cancellation, and a plausible hydrogen-bond explanation must be supported by the candidate's spatial arrangement. Structural changes also entail electronic changes, and the reviewed explanations suggest that models tend to rely on local electronic heuristics without adequately accounting for the accompanying spatial rearrangements.

\subsection{Geometry and Heuristic Diagnostics}
\label{app:geometry-diagnostics}
These analyses use the full 136-task final cohort. Geometry descriptors were computed without loading model predictions. The geometry analyses use source-matched structures that pass candidate-property checks.

\paragraph{Shape and contact contrasts.}
Shape contrast is the maximum candidate-pair retained-heavy-scaffold RMSD after proper-rotation Kabsch alignment, minimized over chirality-preserving scaffold automorphisms. Median splits are 0.52281~\AA\ for dipole/UV--Vis and 0.50689~\AA\ for NMR. Astra's lower/upper-half error counts are 18/59 versus 14/58 for dipole, 13/59 versus 11/58 for UV--Vis, 26/66 versus 21/66 for atomwise $^1$H, and 23/66 versus 21/66 for atomwise $^{13}$C. Larger scaffold-shape contrast therefore does not consistently imply more errors.

A contact screen uses retained N/O-bound H, RDKit-defined N/O acceptors, donor--acceptor graph distance of at least three bonds, H$\cdots$A distance at most 2.5~\AA, and D--H$\cdots$A angle at least $120^\circ$. Contact presence differs across candidates in nine NMR tasks. Atomwise-$^1$H error counts on contrast/noncontrast tasks are 5/9 versus 67/123 (Luna), 4/9 versus 53/123 (Sol), 6/9 versus 41/123 (Astra), and 3/9 versus 47/123 (Opus). Astra atomwise-$^{13}$C gives 6/9 versus 38/123 on the same tasks. With a $110^\circ$ cutoff Astra atomwise-$^1$H gives 9/21 versus 38/111; with $130^\circ$, 3/4 versus 44/128. Distance cutoffs 2.3--2.7~\AA\ do not change membership here.

Per-model/view linear-probability analyses adjust for log minimum target--distractor property distance, candidate count, retained heavy-atom count, and correctness under $P_0$; 2,000 task-bootstrap draws use seed 20260916. The atomwise-$^1$H contact coefficient is 0.1016 per descriptor standard deviation, with a nominal 95\% interval of [0.0212, 0.1827] from 1,999 full-rank bootstrap fits.

\paragraph{Matched carbonyl/reduced pairs.}
Deleting one terminal carbonyl oxygen and sanitizing the reduced graph must reproduce the within-task comparison candidate's non-stereochemical graph. This rule yields 35 pairs in 35 tasks, ten with lower dipole in the carbonyl-rich candidate; every reversal exceeds 0.25~D. Under dipole-magnitude source input, reversal/nonreversal error counts are 7/10 versus 8/25 (Luna), 7/10 versus 7/25 (Sol), 7/10 versus 5/25 (Astra), and 8/10 versus 7/25 (Opus). Restricting to targets inside the pair leaves nine reversal and seventeen nonreversal tasks: errors are 7/9 versus 3/17, 7/9 versus 4/17, 7/9 versus 2/17, and 8/9 versus 2/17, respectively. These selected substitution contrasts challenge a monotonic polarity heuristic.

\phantomsection\label{app:task-dipole-49e99}
Task \hyperref[app:family-49e99a99315e5e96]{49E99A99315E5E96}. Candidate 0, \path{N#CCn1cnnc1}, has dipole 5.37590~D; target candidate 1, \path{N#CC(=O)n1cnnc1}, has 1.45724~D. All four models select candidate 0 under norm input. Astra argues that the additional carbonyl makes the low magnitude less consistent, contradicting the actual ordering. Only Astra switches to the target under vector input, while computing its magnitude and acknowledging frame dependence.

\paragraph{Norm/vector diagnostic.}
On 136 matched tasks, norm/vector correct counts are 89/85 (Luna), 86/94 (Sol), 94/97 (Astra), and 82/84 (Opus). Wrong-to-correct/correct-to-wrong counts are 11/15, 12/4, 7/4, and 13/11; invalid-inclusive exact McNemar p-values are 0.557, 0.077, 0.549, and 0.839. Opus has one invalid vector response and no invalid norm response in this cohort; its valid matched pairs give 82/84 correct and 13/11 gains/losses. Vector norms match scalar inputs within $10^{-10}$~D. In the ten reviewed reversal cases Astra reduces vectors to magnitude and notes frame dependence.

\section{Detailed NMR Properties and Case Analysis}
\label{app:nmr-details}

\paragraph{NMR: restricted observations and errors in interpreting available property information.}
The task references below link to their source molecules, candidate lists, and scaffold drawings in Appendix~\ref{app:task-families}.
Across several models, the supplied NMR representations yield lower source recovery than other informative property channels. For example, Astra obtains 69.9\% with $^1$H peaks and 66.2\% with $^{13}$C peaks, compared with 80.1\% with UV--Vis; Sol obtains 54.4\% for both NMR views, compared with 72.8\% with UV--Vis. This ordering is not universal: Opus performs better with $^{13}$C peaks than with UV--Vis, and Astra's $^1$H result is similar to its HOMO--LUMO-gap result. The relevant question is therefore why the particular NMR property information supplied by this benchmark leaves substantial room for error, rather than whether NMR is intrinsically less informative than other molecular properties.

An important feature of the task is that the NMR observation is restricted to nuclei retained on the visible scaffold (Section~\ref{sec:nmr-scope}). Nuclei within the masked group, including its attached hydrogens, are excluded. The model consequently has to infer a completion from its effect on the shielding of the remaining nuclei, without direct access to that group's own NMR signals. This restriction differs from the whole-molecule scope of the other property channels. For example, if a carbonyl carbon belongs to the masked group, its own $^{13}$C shielding is unavailable even when the candidate structure contains a carbonyl. The observation also consists of calculated gas-phase isotropic shieldings~\citep{gupta2021qm9nmr}, with no experimental multiplet patterns, $J$-couplings, or atom assignments. Applying a familiar whole-spectrum recognition strategy therefore requires care: the model must determine which expected signals are actually within the supplied scope.

Removing hidden-group nuclei eliminates potential discriminative cues, but it does not necessarily eliminate all information about the completion. A concrete example of a difficult residual distinction is Task \hyperref[app:family-79cd04c3dfc1c559]{79CD04C3DFC1C559}.\footnote{The five NMR tasks discussed here belong to an outcome-stratified packet of 24 task--view cases: three cases from each of four strata (Astra alone correct, all four correct, Astra incorrect but another model correct, and all four incorrect), separately for $^1$H and $^{13}$C peaks-only inputs. Selection within strata used SHA256 ordering with a fixed seed.} The retained scaffold is \path{[*:1]C(=O)NC1CCC1}, and the candidates are a methyl carbamate, \path{COC(=O)NC1CCC1}, and a urea, \path{NC(=O)NC1CCC1}. The methoxy carbon of the carbamate lies outside the observed scaffold. The remaining $^{13}$C peak lists are numerically close: the retained-carbonyl shieldings are 28.2095 and 28.0280~ppm, and the Wasserstein-$1$ ($W_1$) distance between the two scaffold distributions is only 0.3562~ppm. All four closed models select the urea when the property information comes from the carbamate. Astra invokes a more deshielded urea carbonyl, while other explanations also rely on general carbonyl-environment arguments. These responses illustrate how broad chemical trends can fail to resolve a fine quantitative distinction after direct substituent signals have been removed. The two computed reference distributions are different.

At the dataset level, candidate separation is associated with model performance. When tasks are grouped using the final cohort's one-third and two-thirds quantiles of the nearest-alternative $W_1$ distance (lower includes the first cutoff; middle includes the second), Astra's $^1$H accuracy increases from 47.8\% to 75.6\% to 86.7\%; its $^{13}$C accuracy increases from 34.8\% to 73.3\% to 91.1\%. These patterns are consistent with numerical proximity contributing to difficulty. Candidate count and chemical composition also differ between groups, however, so the relationship is descriptive. Moreover, all 136 targets have distinct reference distributions at zero tolerance in each NMR representation.

The submitted explanations identify additional problems that cannot be reduced to missing signals. Models sometimes misinterpret the supplied quantity or reason about nuclei that the observation explicitly excludes. Task \hyperref[app:family-e1413048e6c74a5a]{E1413048E6C74A5A} is particularly instructive. All four models correctly select the ester \path{CC(C)COC(=O)CO}, but Luna cites a ``strong downfield 13C shielding peak at 167.54 ppm'' as support for an ester-carbonyl assignment. The carbonyl is inside the masked group and is absent from the observed list. The explanation also applies a carbonyl-region chemical-shift interpretation to an absolute shielding value. For scale, using the matching calculated gas-phase TMS shielding of 186.9704~ppm gives an approximate referenced shift of $186.9704 - 167.54025 \approx 19.43$~ppm, rather than 167.54~ppm.\footnote{The matching calculated gas-phase $^{13}$C TMS reference is tabulated with the QM9NMR dataset at \url{https://moldis-group.github.io/qm9nmr/}. This conversion illustrates the shielding/shift distinction; it is not an experimental peak assignment.} In contrast, Astra, Sol, and Opus discuss the two relatively low-shielding environments of retained oxygen-bound carbons. Because the final choice is correct, this example does not itself contribute to the accuracy deficit; it shows that source recovery can conceal an incorrect account of the property information. The ester/ether comparison below provides a corresponding case in which a scope error accompanies an incorrect choice.

The case records document errors in interpreting the physical meaning of supplied shieldings, their association with local environments, and their observation boundary.

\paragraph{Matched property information and local structural comparisons.}
Astra's advantage is clearest for $^1$H peak lists. It reaches 69.9\% source recovery, compared with 54.4\% for Sol, 55.1\% for Opus, and 37.5\% for Luna. Its advantage over Sol comprises 31 tasks solved only by Astra and ten solved only by Sol, a net difference of 15.4 pp with a descriptive paired 95\% bootstrap interval of [6.6, 24.3].\footnote{This additional NMR interval uses 10,000 task-level bootstrap resamples with seed 20260917. It is post-hoc and pointwise, conditional on the saved runs, without multiple-comparison correction or repeated-generation uncertainty.} Property substitutions provide additional support for a difference in property use. Among the source tasks initially solved correctly by both Astra and Sol and with valid CF substitutions, Astra selects the corresponding CF candidate in 76.6\% of source-averaged outcomes and Sol in 56.0\%. CF outcomes are averaged within each source before averaging across the common source cohort. Furthermore, all candidates have the same peak count in 94.9\% of tasks; within this subset, Astra still obtains 68.2\% source recovery, compared with 52.7\% for Sol and 53.5\% for Opus. The advantage therefore persists when simple peak-count discrimination is unavailable.

\phantomsection\label{app:task-nmr-97ccf}
Task \hyperref[app:family-97ccf0b132ed0cc7]{97CCF0B132ED0CC7} illustrates how an accurate account of candidate connectivity and observation scope can support Astra's decision. The candidates are the ether \path{CO[C@@H](C)C(C)C} and the ester \path{COC(=O)[C@@H](C)C(C)C}, with the ester as the source. The retained methine is directly bonded to oxygen in the ether, but adjacent to a carbonyl carbon in the ester. The supplied $^{13}$C shieldings range from 134.3052 to 168.7333~ppm. Astra selects the ester, explaining that the ether would more strongly deshield the directly oxygen-bound methine and explicitly noting that the masked carbonyl carbon is excluded from the observation. Luna and Opus instead select the ether and invoke the absent carbonyl signal or its expected contribution to the peak pattern. Sol also selects the ether, interpreting the two lowest supplied shieldings as two directly oxygen-bound retained carbons. The decisive distinction is therefore not merely recognizing that an ester contains a carbonyl; it is relating the candidate-specific bonding of the retained atoms to the property information that can actually be observed.

The same case also supplies a behavioral check on Astra's explanation. When the input is replaced with the ether candidate's scaffold property information, which includes shieldings at 101.2638 and 130.2122~ppm, Astra switches to the ether and relates the more strongly deshielded pattern to the oxygen-bound methine and methoxy carbon. Among the four models, Astra alone selects the property-providing candidate in both conditions: Luna reverses the two assignments, while Sol and Opus select the ether under both inputs. These are separate calls with substituted property information, not a conversational correction. This paired outcome strengthens the interpretation that Astra is responding to a structurally relevant change in the supplied property information.

Task \hyperref[app:family-9fcbd2d389e1b5fa]{9FCBD2D389E1B5FA} provides a complementary $^1$H example. The source is \path{CO[C@@H](C=O)C1CC1}, and the alternative is \path{CC(=O)[C@@H](C=O)C1CC1}. Both retain the aldehyde and cyclopropyl portions of the scaffold, while the retained methyl is oxygen-bound in the source and carbonyl-adjacent in the alternative. Astra associates the observed peak at 28.2110~ppm with the oxygen-bound methyl and selects the source. Opus assigns this peak to a methine and selects the carbonyl-linked alternative; Luna associates the much lower 21.5655~ppm peak with deshielding of the chiral proton, whereas Astra identifies an aldehydic contribution. An independent check using the existing counted representation shows that the 28.2110~ppm peak represents three retained hydrogens, contradicting Opus's methine assignment. These counts were not supplied in the peaks-only condition and are used here only to evaluate the explanation. When the alternative candidate's property information is substituted, Astra switches to that candidate and relates the methyl/methine region near 29--30~ppm to the carbonyl-adjacent environments. Again, Astra alone selects the property-providing candidate in both conditions among the four models. Candidate reference values, $W_1$ distances, and the integration check used in this discussion are evaluator-side information, not a candidate-property lookup supplied to the models.

These cases suggest that Astra's stronger $^1$H performance involves more accurate coordination of NMR interpretation with local structural context. Its successful explanations identify which atoms remain observable, distinguish direct oxygen attachment from carbonyl adjacency, and associate numerical patterns with candidate-specific environments. These outputs provide checkable correspondences between the supplied property information, candidate structures, and stated explanations. Sol and Opus correctly use related distinctions in other cases. For example, in Task \hyperref[app:family-894bbf79c573c29a]{894BBF79C573C29A}, they correctly select an isopropyl carbamate, while Astra selects the amide alternative and gives a different NH/methine assignment. The overall model ranking is also nucleus-dependent: Opus exceeds Astra in the $^{13}$C peaks-only and atomwise conditions. The aggregate results establish higher $^1$H peaks-only source recovery for Astra, while the selected cases document both property-consistent local comparisons and explanation errors.

\section{Complete Task Families}
\label{app:task-families}
This section lists all 136 source tasks and their 394 candidate entries. For each task, we provide the source-molecule SMILES, the masked scaffold as SMILES and a structural drawing, and the complete candidate list in its original order. Tasks are identified by the same 16-character IDs used throughout the paper; task references link directly to the corresponding entries below. Candidate indices are zero-based, as in the task inputs; a superscript asterisk marks the source candidate. In the scaffold drawings, $*$ denotes the masked fragment. SMILES strings, including stereochemical and charge annotations, are reproduced unchanged from the task records.

\begingroup
\footnotesize
\setlength{\tabcolsep}{4pt}
\renewcommand{\arraystretch}{1.1}
\begin{longtable}{@{}p{0.40\linewidth}p{\dimexpr0.36\linewidth-2\tabcolsep\relax}p{\dimexpr0.24\linewidth-2\tabcolsep\relax}@{}}
\caption{Complete task families: source molecules, masked scaffolds, and candidate SMILES.}\label{tab:complete-task-families}\\
\toprule
\textbf{Task and SMILES} & \textbf{Candidate list} & \textbf{Scaffold} \\
\midrule
\endfirsthead
\multicolumn{3}{l}{\small\textit{Complete task families (continued)}}\\
\toprule
\textbf{Task and SMILES} & \textbf{Candidate list} & \textbf{Scaffold} \\
\midrule
\endhead
\midrule
\multicolumn{3}{r}{\small\textit{Continued on next page}}\\
\endfoot
\bottomrule
\endlastfoot
\begin{minipage}[c]{\linewidth}\raggedright
\phantomsection\label{app:family-005a3b23ab24ffb0}\textbf{Task 005A3B23AB24FFB0}\\
Source: \texttt{O=CCCCC=O}\\
Scaffold: \texttt{[*:1](CC=O)CC=O}\strut
\end{minipage} &
\begin{minipage}[c]{\linewidth}\raggedright
0$^{*}$: \texttt{O=CCCCC=O}\\
1: \texttt{O=CCOC(=O)CC=O}\strut
\end{minipage} &
\begin{minipage}[c]{\linewidth}\centering
\includegraphics[width=\linewidth]{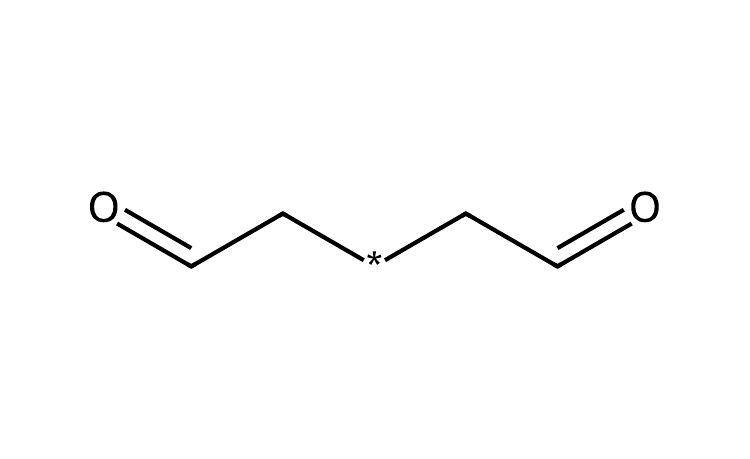}
\end{minipage}\\[5pt]
\midrule
\begin{minipage}[c]{\linewidth}\raggedright
\phantomsection\label{app:family-0293945fd5430b3d}\textbf{Task 0293945FD5430B3D}\\
Source: \texttt{CNC(=O)OC1CC1}\\
Scaffold: \texttt{[*:1](C)C(=O)OC1CC1}\strut
\end{minipage} &
\begin{minipage}[c]{\linewidth}\raggedright
0: \texttt{CC(=O)C(=O)OC1CC1}\\
1$^{*}$: \texttt{CNC(=O)OC1CC1}\strut
\end{minipage} &
\begin{minipage}[c]{\linewidth}\centering
\includegraphics[width=\linewidth]{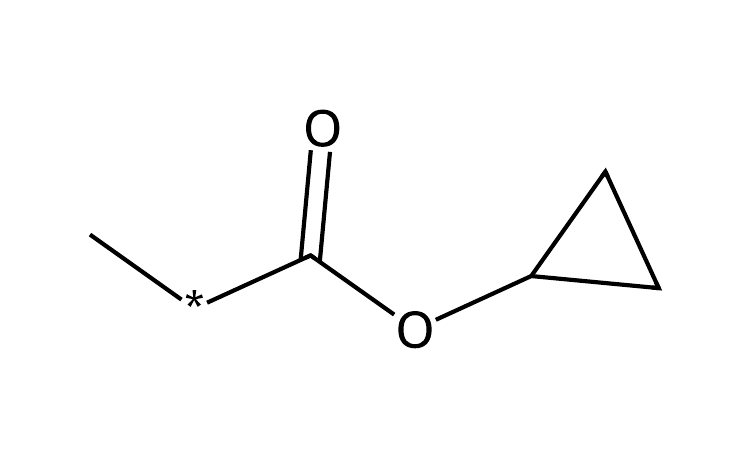}
\end{minipage}\\[5pt]
\midrule
\begin{minipage}[c]{\linewidth}\raggedright
\phantomsection\label{app:family-043694cc3a98d45d}\textbf{Task 043694CC3A98D45D}\\
Source: \texttt{CC[C@H]1OC1(C)C}\\
Scaffold: \texttt{[*:1]C[C@H]1OC1(C)C}\strut
\end{minipage} &
\begin{minipage}[c]{\linewidth}\raggedright
0: \texttt{COC[C@H]1OC1(C)C}\\
1: \texttt{CC(=O)C[C@H]1OC1(C)C}\\
2$^{*}$: \texttt{CC[C@H]1OC1(C)C}\strut
\end{minipage} &
\begin{minipage}[c]{\linewidth}\centering
\includegraphics[width=\linewidth]{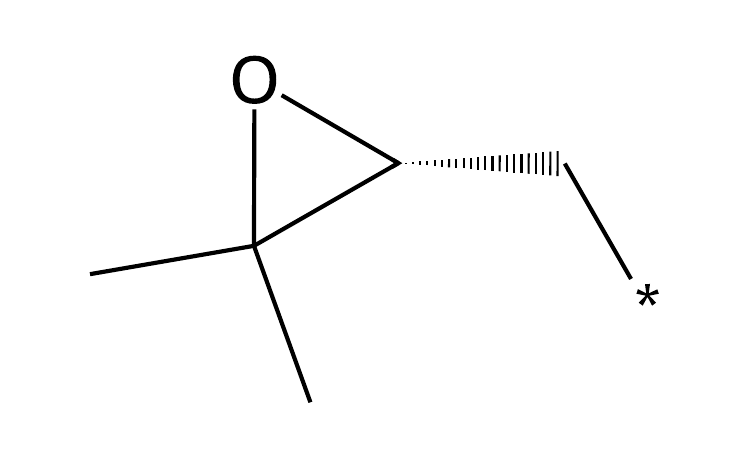}
\end{minipage}\\
\end{longtable}
\endgroup

\stopcontents[appendices]
\end{document}